\documentclass[11pt,superscriptaddress,aps,prd,preprint,showpacs]{revtex4}
\usepackage[dvips]{graphicx}
\usepackage{amssymb}
\usepackage{amsmath}
\usepackage{graphicx}
\usepackage{hyperref}
\usepackage{color}

\newcommand{\sla}[1]{\hbox{{$#1$}\llap{$/$}}}

\usepackage{amsfonts}
\usepackage{slashed}
\newcommand{\bea}{\begin{eqnarray}}
\newcommand{\eea}{\end{eqnarray}}
\newcommand{\pa}{\partial}

\def\ks{k\!\!\!/}

\def\ps{p\!\!\!/}

\def\us{u\!\!\!/}
\def\vs{v\!\!\!/}

\def\ds{\partial\!\!\!/}

\begin{document}

\title{On quantum aspects of the higher-derivative Lorentz-breaking extension of QED}

\author{T. Mariz}
\affiliation{Instituto de F\'\i sica, Universidade Federal de Alagoas, 57072-270, Macei\'o, Alagoas, Brazil}
\email{tmariz@fis.ufal.br}

\author{J. R. Nascimento}
\affiliation{Departamento de F\'{\i}sica, Universidade Federal da Para\'{\i}ba, Caixa Postal 5008, 58051-970, Jo\~ao Pessoa, Para\'{\i}ba, Brazil}
\email{jroberto, petrov@fisica.ufpb.br}

\author{A. Yu. Petrov}
\affiliation{Departamento de F\'{\i}sica, Universidade Federal da Para\'{\i}ba, Caixa Postal 5008, 58051-970, Jo\~ao Pessoa, Para\'{\i}ba, Brazil}
\email{jroberto, petrov@fisica.ufpb.br}

\author{C. Marat Reyes}
\affiliation{Departamento de Ciencias B\'{a}sicas, Universidad del B\'{\i}o B\'{\i}o,\\ 
Casilla 447, Chill\'{a}n, Chile}
\email{creyes@ubiobio.cl}


\begin{abstract}
We consider the higher-derivative Lorentz-breaking extension of QED, where the new terms are the Myers-Pospelov-like ones in gauge and spinor sectors, and the higher--derivative CFJ term. For this theory,
 we study its tree-level dynamics, discuss the dispersion relation, and present one more scheme for its perturbative generation, including the finite temperature 
 case.  Also, we develop a method to study perturbative 
unitarity based on consistent rotation of the theory to Euclidean space. We use this method to verify explicitly that 
 for special choices of the Lorentz-breaking vector, the unitarity is preserved at the one-loop level, even in the presence of higher time derivatives.

\end{abstract}


\maketitle

\section{Introduction}

Formulation of the Lorentz-breaking extension of the standard model called attention to studies of Lorentz-breaking extensions for many field theory models, and, first of all, for QED \cite{ColKost}. Conclusions obtained by treating different aspects of various extensions of QED in dozens of papers became paradigmatic results for Lorentz-breaking theories in general. Among the most important directions of their study, one can emphasize searches of exact solutions, canonical quantization and calculations of quantum corrections. These studies have allowed to put strong bounds on Lorentz violating quantum field theory
models \cite{datatables}. 
Within this context, an important role is naturally played by higher-derivative Lorentz-breaking extensions of QED. Indeed, it is well known that an effective action is nonlocal and can be represented in the form of the derivative expansion. Moreover, the higher-derivative terms naturally emerge within the string context \cite{HDString}. Therefore, one naturally faces a problem of studying different issues related to higher-derivative Lorentz-breaking extensions of QED.  The first step in such study has been carried out in \cite{Myers:2003fd} where the so-called Myers-Pospelov (MP) term, that is, the first higher-derivative Lorentz-breaking term in QED, has been proposed. This term attracted a great interest due to the fact that a special choice of the Lorentz-breaking vector allows to rule out the higher time derivatives from this term, thus avoiding unitarity breaking which is known to be the main problem of higher-derivative theories. A number of studies of unitarity issues for QED with the additive MP term have been performed in \cite{CMR}. Some other tree-level results for this theory can be found in \cite{Ganguly}, and its phenomenological applications -- in \cite{BP}. Further, the higher-derivative terms were shown to arise as quantum corrections, first, for the case when the Lorentz symmetry breaking is introduced through the third-rank constant tensor \cite{TM} (which for a certain choice of this tensor yields the higher-derivative CFJ-like term discussed in \cite{KostMew0,Leite:2013pca}), second, for the case where the Lorentz symmetry breaking is introduced through a constant vector, with the nonminimal coupling is present \cite{MNP}. It was shown that in these cases the resulting higher-derivative terms are finite. Therefore, one can naturally establish the questions, first, about other possible schemes allowing to generate the higher-derivative Lorentz-breaking terms, second, about the tree-level behavior of the QED with additive higher-derivative Lorentz-breaking terms, which clearly would modify propagators and ultraviolet behavior of the theory.  In this paper, we address namely these questions. To be more precise, in this paper we introduce the higher-derivative terms in the gauge sector and discuss the impacts of higher derivatives for the propagator and unitarity.

The structure of this paper looks like follows. In the section \ref{II}, we introduce the classical action of the 
gauge sector of the Lorentz-breaking extended QED with higher derivatives. In the section 
\ref{III}, we carry out the one-loop calculation of the higher-derivative Lorentz-breaking terms in the gauge sector with use of the new coupling, both at zero 
and finite temperature. In the section \ref{IV} we discuss the related unitarity issues and explicitly demonstrate that even in the presence 
of the higher time derivatives, unitarity is preserved. Finally, in the section \ref{V}, we summarize our results.

\section{Classical action and dispersion relations}\label{II}
 
Let us consider the higher-derivative (HD) extension of QED looking like
\bea
\label{free}
{\cal L}_{HD}=-\frac{1}{4}F_{\mu\nu}F^{\mu\nu}-\frac{1}{M}\epsilon^{\beta\mu\nu\lambda}u_{\beta}A_{\mu}\Big(c_1(u\cdot \pa)^2-c_2u^2\Box \big)F_{\nu\lambda}.
\eea
Here $u_{\mu}$ is a dimensionless vector, $M$ is a mass scale, which is typically suggested to be of the order of the Planck mass \cite{Myers:2003fd}, $c_1$ and $c_2$ are some dimensionless numbers. They accompany the Myers-Pospelov \cite{Myers:2003fd} and higher-derivative CFJ-like \cite{Leite:2013pca} terms respectively (we note that within many schemes these terms arise simultaneously, see f.e. \cite{MNP}). We note that both these terms are CPT-odd, and they represent themselves as specific particular examples of higher-derivative Lorentz-breaking extensions of the gauge sector discussed in details in \cite{KostMew1}.

Since this theory is gauge invariant, we can impose the usual Feynman gauge which does not affect the higher-derivative terms. The resulting quadratic Lagrangian for the essentially transversal $A_{\mu}$ will be given by the expression 
\bea
\label{quad}
{\cal L}=\frac{1}{2}A_{\mu}\Delta^{\mu\nu}A_{\nu},
\eea
with
\bea\label{Delta}
\Delta^{\mu\lambda}=\Box\eta^{\mu\lambda}+\frac{4}{M}\Sigma\epsilon^{\beta\mu\nu\lambda}u_{\beta}\pa_{\nu},
\eea
where we introduced the notation $\Sigma=c_2u^2\Box-c_1(u\cdot\pa)^2$. 
As a result, one will have just the propagator, whose explicit form is
\bea
\label{f1a}
G_{\nu\lambda}(x-x')=\left[
A_1\eta_{\nu\lambda}+A_2u_{\nu}u_{\lambda}+A_3u_{\nu}\pa_{\lambda}+A_4u_{\lambda}\pa_{\nu}+A_5\pa_{\nu}\pa_{\lambda}+A_6\epsilon_{\nu\lambda\rho\sigma}u^{\sigma}\pa^{\rho}
\right]\delta^4(x-x').
\eea
Defining
\bea
\label{f2a}
D&=&u^2\Box-(u\cdot \pa)^2,\quad\, Q=\Box^2-\frac{16\Sigma^2}{M^2}D,
\eea
we get
\bea
\label{f3a}
A_1&=&\frac{\Box}{Q},\quad\, A_2=-\frac{16\Sigma^2}{M^2Q},\nonumber\\
A_3&=&A_4=\frac{16\Sigma^2(u\cdot\pa)}{M^2Q\Box},
\nonumber\\
A_5&=&\frac{4\Sigma A_6u^2}{M\Box}=-\frac{16\Sigma^2u^2}{M^2Q\Box},\quad\, A_6=-\frac{4\Sigma}{MQ}.
\eea
Throughout this paper, we are using the definition of the Levi Civita tensor $\epsilon^{0123}= -\epsilon_{0123}=1$. 

In momentum space we write 
\bea
G_{\nu\lambda}(p)&=&\frac{1}{Q(p)} \left[
-p^2\eta_{\nu\lambda}-a^2 \left(u_{\nu}u_{\lambda}-\frac{(u\cdot p) }{p^2} 
(u_{\nu}p_{\lambda}+u_{\lambda}p_{\nu})+\frac{u^2}{p^2}p_{\nu}p_{\lambda}\right)  +a i \epsilon_{\nu\lambda\rho\sigma}u^{\sigma}p^{\rho}
\right].
\eea
where 
\bea
a=\frac{4\Sigma(p)}{M} \,,
\eea
and $Q(p)$ and $\Sigma(p)$ are just the momentum space counterparts of the same expressions. That is,
\bea
Q(p)=(p^2)^2-a^2D(p)\,,
\eea
with
\bea \label{D}
D(p)=(u\cdot p)^2-u^2p^2\,.
\eea
We note that this propagator involves the contributions asymptotically behaving like $\frac{1}{\Box}$, which indicates that the UV behavior is the same as in usual theories without higher derivatives (for example, the term $A_1\eta_{\nu\lambda}$ asymptotically behaves as $\frac{1}{k^2}$), and renormalization properties will not be improved compared with the usual QED. The similar situation occurs in three-dimensional QED with higher-derivative CFJ term $\kappa\epsilon^{\mu\nu\lambda}A_{\mu}\pa_{\nu}\Box A_{\lambda}$, where one has
$$
[\Box(\eta^{\mu\nu}+\kappa\epsilon^{\mu\nu\lambda}\pa_{\lambda})]^{-1}=
\frac{\eta_{\nu\rho}}{\Box(1+\kappa^2\Box)}
+\frac{\kappa^2\pa_{\nu}\pa_{\rho}}{\Box(1+\kappa^2\Box)}-\frac{\kappa\epsilon_{\nu\rho\sigma}\pa^{\sigma}}{\Box(1+\kappa^2\Box)}.
$$
Here, the term proportional to $\pa_{\nu}\pa_{\rho}$ asymptotically behaves as $k^{-2}$, thus the UV asymptotics is the same as in the usual case.

To find the dispersion relations, one should consider the denominators of (\ref{f3a}) and carry out the Fourier transform, so, 
from the denominator $Q$ one finds the unusual dispersion relation (where $u^2=u^2_0-\vec{u}^2$ is the usual square of the vector $u^{\mu}$ in Minkowski space), whose some aspects have been earlier studied also in \cite{Campanelli}:
\bea
(E^2-\vec{p}^2)^2+\frac{16}{M^2}\Big(c_2u^2(E^2-\vec{p}^2)-c_1(u_0E-\vec{u}\cdot\vec{p})^2
\Big)^2\Big(u^2(E^2-\vec{p}^2)-(u_0E-\vec{u}\cdot\vec{p})^2
\Big)=0.
\eea
This dispersion relation in general cannot be simplified since there is no fundamental reason to impose the relation $c_1=c_2$ forever. Here we emphasize some typical situations.

1. The vector $u^{\mu}$ is light-like, $u_{\mu}u^{\mu}=0$. In this case the CFJ-like term vanishes (the same situation is observed if $c_2=0$), and we have the simplified dispersion relation:
\bea
(E^2-\vec{p}^2)^2-\frac{16c^2_1}{M^2}(u_0E-\vec{u}\cdot\vec{p})^6=0.
\eea

2. For $c_1=c_2$, 
we have the following simplification of the dispersion relation:
\bea
(E^2-\vec{p}^2)^2+\frac{16c^2_1}{M^2}\Big(u^2(E^2-\vec{p}^2)-(u_0E-\vec{u}\cdot\vec{p})^2
\Big)^3=0.
\eea

3. The vector $u^{\mu}$ is space-like, with $u_0=0$, and $c_2=0$ (no CFJ-like term). In this case 
we can avoid the presence of higher time derivatives (so, the theory does not involve ghosts, being hence most probably unitary), and
\bea
(E^2-\vec{p}^2)^2-\frac{16c^2_1}{M^2}(\vec{u}\cdot\vec{p})^4
\Big(\vec{u}^2(E^2-\vec{p}^2)+(\vec{u}\cdot\vec{p})^2
\Big)=0.
\eea

4. If $u^{\mu}$ is time-like and has only $u_0$ non-zero component, with $u_i=0$, we also have the absence of higher time derivatives (so,  unitarity is again most probably achieved).

To study unitarity in our theory we must determine
the physical degrees of freedom of the gauge field and the correct 
 $i\epsilon$ prescription in order to perform a consistent Wick rotation to Euclidean space, as we explain in section \ref{IV}.
 
Let us begin to study the extra conditions on the gauge field, arising through contracting $\partial_{\mu}$ and $u_{\mu}$ with $\Delta^{\mu \lambda}$ in Eq. \eqref{Delta}. We obtain $(\partial \cdot A)=(u \cdot A)=0$,
which indicates that we must express the gauge field in terms of 
polarization vectors perpendicular to $p$ and $u$. The strategy to obtain these polarization vectors is to start with 
two real transverse vectors
$e^{(a)}_{\mu}$, with $a=1,2$, and then 
change to transverse complex ones $\varepsilon^{(\lambda)}_{\mu}$, with $\lambda=\pm $.

Let us consider two linear polarization vectors $e_{\mu}^{(a)}$,
satisfying the relation
\begin{eqnarray}
e_{\mu\nu}=-\sum_{a=1,2} e_{\mu}^{(a)}e_{\nu}^{(b)}\,,
\end{eqnarray}
and 
\begin{eqnarray}
\eta^{\mu \nu} e_{\mu}^{(a)}e_{\nu}^{(b)}=-\delta^{ab}.
\end{eqnarray}
Now we introduce the projector $P^{(\lambda)}_{\mu \nu}$
\bea\label{Plambda}
P^{(\lambda)}_{\mu \nu}=\frac{1}{2}(e_{\mu \nu }+i \lambda \epsilon  _{\mu \nu})\,,
\eea
which projects any four vector $v^{\mu}$
onto the hyperplane orthogonal to $u^{\nu}$ and $p^{\lambda}$ vectors, with 
\begin{eqnarray}\label{rele1}
e^{\mu\nu}&=&\eta^{\mu \nu}-\frac{(u\cdot p)}{D}
(u^{\mu}p^{\nu}+u^{\nu}p^{\mu})+\frac{p^2}{D} u^{\mu}u^{\nu}
+\frac{u^2}{D} p^{\mu}p^{\nu}\,,
\\ \label{rele2}
\epsilon^{\mu \nu}&=&\frac{\epsilon^{\mu \lambda \rho \nu}u_{\lambda}p_{\rho}}{\sqrt{D}}\,.
\end{eqnarray}
Indeed, one can show that these tensors are orthogonal to $p$ and $u$, i.e.,
\bea 
e^{\mu \nu} u_{\nu}=e^{\mu \nu} p_{\nu}=0\,, \nonumber\\
\epsilon^{\mu \nu} u_{\nu}=\epsilon^{\mu \nu} p_{\nu}=0\,.
\eea
They also satisfy the relations
\bea 
e^{\mu \nu} e_{\nu}^{\; \beta}=e^{\mu \beta} \,,
\\ e^{\mu \nu}  \epsilon_{\nu}^{\; \beta}=
\epsilon^{\mu \nu}  e_{\nu}^{\; \beta}=\epsilon^{\mu \beta} \,,\nonumber
\\  \epsilon^{\mu \nu} 
 \epsilon_{\nu}^{\; \beta}=-e^{\mu \beta}\,.\nonumber
\eea
Using these properties, one can show that these tensors diagonalize the equation of motion or 
$\Delta_{\mu \nu}$ in 
Eq. \eqref{Delta}, since
\bea 
P^{(\lambda)}_{\mu \nu} \eta^{\nu \alpha}  P^{(\lambda')}_{\alpha \beta}
=\delta^{\lambda \lambda'} P^{(\lambda)}_{\mu \beta}\,,
\\
P^{(\lambda)}_{\mu \nu} \epsilon^{\nu \alpha}  P^{(\lambda')}_{\alpha \beta}
=(-i\lambda)\delta^{\lambda \lambda'} P^{(\lambda)}_{\mu \beta} \,.
\eea
We can define the analogues to the circular polarization vectors
\begin{eqnarray}
\varepsilon^{(+)}_{\mu}=\frac{1}{2}(e_{\mu}^{(1)}+ie_{\mu}^{(2)}),
\nonumber\\
\varepsilon^{(-)}_{\nu}=\frac{1}{2}(e_{\nu}^{(1)}+ie_{\nu}^{(2)}),
\end{eqnarray}
such that 
\begin{eqnarray}\label{proj}
P^{(\lambda)}_{\mu\nu}=-  \varepsilon^{(\lambda)}_{\mu}({p})  \varepsilon^{*(\lambda)}_{\nu} ({p})\,.
\end{eqnarray}
The transverse propagator is
 \begin{eqnarray}\label{PROPAGATOR}
iG^T_{ \mu \nu }(p)&=&\sum_{\lambda=\pm}   \left(  \frac{P^{(\lambda)}_{\mu \nu}  }{p^2+\lambda  a  \sqrt{D}   }  \right)_{p^2\to p^2+i\epsilon}\,,
\end{eqnarray}
with
\bea\label{Da}
a=\frac{4(c_1(u\cdot p)^2-c_2u^2p^2)}{M}\,,
\eea
where we have included the usual prescription $p^2\to p^2+i\epsilon$ to fix the position of the poles in the complex energy plane. 
This prescription gives the correct position of the poles in the second and fourth quadrant when considering the case $u$
purely spacelike or taking the limit $M\to \infty$. We use this propagator in the Section \ref{IV}. 

\section{Perturbative generation}\label{III}

Let us consider the perturbative generation of the Myers-Pospelov term.
One scheme, based on the magnetic coupling, has been developed in \cite{MNP}, where it was shown to yield the finite corrections. Here we deal with another one.
Let us consider the following fermionic Lagrangian \cite{Myers:2003fd}, representing itself as an example of a family of Lorentz-breaking higher-derivative fermionic Lagrangians considered in \cite{KostMew2}:
\begin{equation}
\label{fermi}
\mathcal{L}_{f} = \bar{\psi}\left(i \slashed{D}-m+\frac{\eta_2}{M} \gamma_5 \slashed{v}(v \cdot D)^2\right)\psi,
\end{equation}
where $D_{\mu}=\pa_{\mu}-ieA_{\mu}$, $v_{\mu}$ is a dimensionless vector, $M$ is the Planck mass (as above), and $\eta_2$ is some dimensionless number. So, we have the following explicit form of the Lagrangian:
\begin{eqnarray}
{\cal L}_f&=&\bar{\psi}\Big(i\slashed{\pa}-m+\frac{\eta_2}{M} \gamma_5 \slashed{v}(v \cdot \pa)^2+e\slashed{A}+\nonumber\\ &+&\frac{\eta_2}{M} \gamma_5 \slashed{v}v^{\mu}v^{\nu}(-ie(A_{\mu}\pa_{\nu}+A_{\nu}\pa_{\mu})-ie(\pa_{\mu}A_{\nu})-e^2A_{\mu}A_{\nu})
\Big)\psi.
\end{eqnarray}
We note that the similar coupling, but including third derivatives, has been used in \cite{NPR}.

\begin{figure}[ht]
\centering
\includegraphics{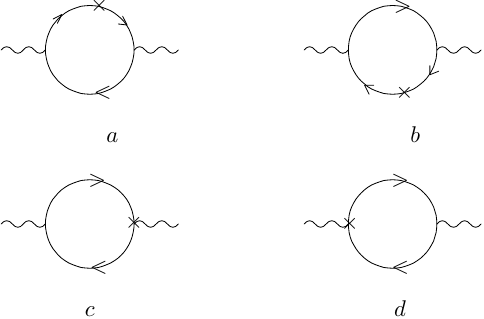}
\caption{\label{Fig1} The contributions to the two-point function of $A^{\mu}$ with triple vertices.}
\end{figure}

 One can easily observe that the one-loop effective action of second order in $A$, in lower order in $\eta_2$, is given by two contributions graphically represented by Fig.~\ref{Fig1}. The first of them involves only usual (minimal) vertices proportional to $e$ which do not involve any $v^{\mu}$ vector, and $\eta_2$ arises from the expansion of the propagator, and is given by graphs $a$, $b$ of Fig.~\ref{Fig1}. It looks like
\begin{equation}
S^{(2)}_{eff} = \frac{ie^2}{2} \int d^{4}x \, \Pi_{v}^{\mu \nu} A_{\mu} A_{\nu},
\end{equation}
where
\begin{equation}
\Pi_{v}^{\mu \nu}= \int \frac{d^{4}p}{(2\pi)^{4}} \mathrm{tr}\,G_v(p)\gamma^\mu G_v(p-k)\gamma^\nu
\end{equation}
with
\begin{equation}
\label{prop}
G_v(p) = \frac{1}{\slashed{p}-m-\frac{\eta_2}{M} \gamma_5 \slashed{v}(v \cdot p)^2}.
\end{equation}

Now, by applying the expansion
\begin{equation}
G_v(p) = S(p)+S(p)\frac{\eta_2}{M} \gamma_5 \slashed{v}(v \cdot p)^2S(p)+\cdots,
\end{equation}
with  $S(p)=(\slashed{p}-m)^{-1}$, we can easily single out the terms of first order in $\eta_2$, by writing
\begin{eqnarray}\label{zaszasz}
\Pi_{1MP}^{\mu \nu}(k)&=& \frac{\eta_2}{M} \mu^{4-D}\int \frac{d^{D}p}{(2\pi)^{D}} \mathrm{tr}[S(p)\gamma_{5}\slashed{v}(v \cdot p)^{2}S(p)\gamma^{\mu}S(p-k)\gamma^\nu \nonumber\\
&&+S(p)\gamma^{\mu}S(p-k)\gamma_{5}\slashed{v}(v \cdot p-v \cdot k)^{2}S(p-k)\gamma^\nu],
\end{eqnarray}
where we have promoted the integral to the $D$-dimensional space-time, so that $d^4p/(2\pi)^4$ is replaced by to $\mu^{4-D}d^Dp/(2\pi)^D$, with $\mu$ being a renormalization scale.

In order to calculate the above integrals, we will use the Feynman parametrization, so that Eq.~(\ref{zaszasz}) is rewritten as
\begin{eqnarray}\label{zaszasz1}
\Pi_{1MP}^{\mu \nu}(k)&=& \frac{\eta_2}{M} \mu^{4-D}\int_0^1 dx \int \frac{d^{D}p}{(2\pi)^{D}} \frac{1}{(p^2-M_x^2)^3}\nonumber\\
&&\times\mathrm{tr}[2(1-x)(\slashed{q}+m)\gamma_{5}\slashed{v}(v \cdot q)^{2}(\slashed{q}+m)\gamma^\mu(\slashed{q}_1+m)\gamma^\nu \nonumber\\
&&+2x(\slashed{q}+m)\gamma^\mu(\slashed{q}_1+m)\gamma_{5}\slashed{v}(v \cdot q_1)^{2}(\slashed{q}_1+m)\gamma^\nu],
\end{eqnarray}
where $M_x^2=m^2-(1-x)x k^2$, $q^\mu=p^\mu+xk^\mu$, and $q_1^\mu=q^\mu-k^\mu$. Then, after the  calculating the trace, we obtain
\begin{equation}
\Pi_{1MP}^{\mu \nu}(k) = \frac{\eta_2}{M} \mu^{4-D}\int_0^1 dx \int \frac{d^{D}p}{(2\pi)^{D}} \frac{1}{(p^2-M_x^2)^3} \sum_{i=1}^4 I_i^{\mu\nu},
\end{equation}
with
\begin{eqnarray}
I_1^{\mu\nu} &=& -16\epsilon^{\mu\alpha\beta\gamma}q_\alpha v_\beta k_\gamma [(1-x)q^{\nu}(v\cdot q)^2+x q_1^{\nu}(v\cdot q_1)^2], \nonumber\\
I_2^{\mu\nu} &=& -16\epsilon^{\alpha\nu\beta\gamma}q_\alpha v_\beta k_\gamma [(1-x)q^{\mu}(v\cdot q)^2+x q_1^{\mu}(v\cdot q_1)^2], \nonumber\\
I_3^{\mu\nu} &=& 8\epsilon^{\mu\nu\beta\gamma}v_\beta k_\gamma [(1-x)(v\cdot q)^2(q^2-m^2)+2x(v\cdot q_1)^2(q\cdot q_1-m^2)], \nonumber\\
I_4^{\mu\nu} &=& 8\epsilon^{\mu\nu\alpha\beta}q_\alpha v_\beta [(1-x)(v\cdot q)^2(q^2+m^2-2 q\cdot q_1)-x(v\cdot q_1)^2(q_1^2+m^2-2 q\cdot q_1)].
\end{eqnarray}
In the following, after we integrate over $d^Dp$ and expand the result around $D=4$, we have
\begin{eqnarray}
\Pi_{1MP}^{\mu \nu} &=& -\frac{i}{4\pi}\frac{\eta_2}{M}\epsilon^{\mu\nu\beta\gamma}v_\beta k_\gamma \int_0^1 dx \left(\frac{1}{\epsilon}-\ln\frac{M_x}{\mu'}\right) \left[\left(10  (1-x)^2 x^2k^2+(1-6 (1-x) x)m^2 \right)v^2\right. \nonumber\\
&&\left.-4 (1-x) (2-5 (1-x) x) x (v\cdot k)^2\right]-
\frac{i}{8\pi}\frac{\eta_2}{M}\epsilon^{\mu\nu\beta\gamma}v_\beta k_\gamma \int_0^1 dx \nonumber\\
&&\times \frac{1}{M_x^2}\left[(1-6(1-x) x)M_x^4v^2-2(1-x)^2x^2k^2 (3-4 (1-x) x) (v\cdot k)^2\right],
\end{eqnarray}
where $\epsilon=4-D$ and $\mu'^2=4\pi\mu^2e^{-\gamma}$. We note that, as $\int_0^1 dx (1-6 (1-x) x)=0$, the divergent contribution does not depend on the mass $m$. 

Finally, after we integrate over the parameter $x$, we obtain
\begin{eqnarray}
\Pi_{1MP}^{\mu \nu} &=& -\frac{1}{12\pi^2\epsilon'} \frac{\eta_2}{M} \left[k^{2}v^{2} - 2 (v \cdot k)^{2}\right] \epsilon^{\mu\nu\lambda\rho}n_{\lambda} k_{\rho} \\
&&+\frac{1}{72 \pi ^2}\frac{\eta_2}{M}\left[5 k^2+12 m^2+\frac{6\left(k^4-2 k^2 m^2-8 m^4\right)}{\sqrt{k^2 \left(4 m^2-k^2\right)}}\csc^{-1}\left(\frac{2m}{\sqrt{k^2}}\right)\right] v^2 \epsilon^{\mu\nu\lambda\rho}v_{\lambda} k_{\rho} \nonumber\\
&&-\frac{1}{18 \pi ^2}\frac{\eta_2}{M}\left[1-\frac{3m^2}{k^2}+\frac{3\left(k^4-2 k^2 m^2+4 m^4\right)}{\sqrt{k^6 \left(4 m^2-k^2\right)}}\csc^{-1}\left(\frac{2m}{\sqrt{k^2}}\right)\right] (v\cdot k)^2 \epsilon^{\mu\nu\lambda\rho}v_{\lambda} k_{\rho}, \nonumber
\end{eqnarray}
with $\frac{1}{\epsilon'}=\frac{1}{\epsilon}-\ln\frac{m}{\mu'}$. Here, we can consider the limits $k^2\ll m^2$ ($m\neq0$) and $k^2\gg m^2$ ($m=0$), so that we get
\begin{equation}\label{km}
\Pi_{1MP}^{\mu \nu} = -\frac{1}{12\pi^2\epsilon'} \frac{\eta_2}{M} \left[k^{2}v^{2} - 2 (v \cdot k)^{2}\right] \epsilon^{\mu\nu\lambda\rho}v_{\lambda} k_{\rho} + {\cal O}\left(\frac{k^2}{m^2}\right)
\end{equation}
and
\begin{eqnarray}
\Pi_{1MP}^{\mu \nu} &=& -\frac{1}{12\pi^2\epsilon''} \frac{\eta_2}{M} \left[k^{2}v^{2} - 2 (v \cdot k)^{2}\right] \epsilon^{\mu\nu\lambda\rho}v_{\lambda} k_{\rho} \nonumber\\
&&+\frac{1}{72\pi^2} \frac{\eta_2}{M}\left[5k^{2}v^{2} - 4(v \cdot k)^{2}\right] \epsilon^{\mu\nu\lambda\rho}v_{\lambda} k_{\rho}+ {\cal O}\left(\frac{m^2}{k^2}\right),
\end{eqnarray}
respectively, where we have also defined $\frac{1}{\epsilon''}=\frac{1}{\epsilon}-\ln\frac{k}{\mu'}$, with $k=\sqrt{k^2}$.

Therefore, for the induced bosonic Myers-Pospelov term from the contribution (\ref{km}), which corresponds to the non-zero mass, we have
\begin{eqnarray}\label{zzfdd}
S_{1MP} =  \frac{e^2}{12\pi^2\epsilon'}  \frac{\eta_2}{M} \int d^{4}x \left[v^{2} v_{\beta} \epsilon^{\beta \mu \nu \lambda} A_{\mu} \Box F_{\nu \lambda} -2 v^{\alpha}F_{\alpha \mu} (v \cdot \partial) v_{\beta} \epsilon^{\beta \mu \nu \lambda} F_{\nu \lambda} \right].
\end{eqnarray}
This enforces the fact that the above higher-derivative terms should be introduced from the very beginning (\ref{free}), so that we have a consistent subtraction of the divergences.


\begin{figure}[ht]
\centering
\includegraphics{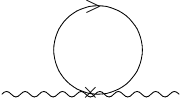}
\caption{\label{Fig2} The contributions to the two-point function of $A^{\mu}$ with the quartic vertex.}
\end{figure}

Then, the quartic vertex  (corresponding Feynman diagram is given by Fig.~\ref{Fig2}) evidently should give a zero contribution. Indeed, this diagram can yield only the Proca-like term $(v\cdot A)^2$ with no derivatives, since there are no derivatives of $A_{\mu}$ in the classical action, and the only relevant graph is a tadpole, so, the integration over the internal momentum cannot yield a contribution depending on the external momentum, and this term is inconsistent with the gauge symmetry (its vanishing can be shown explicitly, as well).

 In order to find the remaining first-order contribution in $\eta_2$ presented by graphs $c$ and $d$ of Fig. 1, we should consider the contraction of two vertices: the first of them is the usual $e\bar{\psi}\slashed{A}\psi$, and the second one is 
$-ie\bar{\psi}\frac{\eta_2}{M} \gamma_5 \slashed{v}v^{\mu}v^{\nu}((A_{\mu}\pa_{\nu}+A_{\nu}\pa_{\mu})+(\pa_{\mu}A_{\nu}))\psi$, where the propagator is the free one (indeed, expanding (\ref{prop}), we will get only the higher-order contributions).
Its explicit form is
\begin{eqnarray}
\label{s2}
S_{2MP}=ie^2\frac{\eta_2}{M}v^{\nu}v^{\mu}A_{\mu}(-k)A_{\lambda}(k)
\int\frac{d^4p}{(2\pi)^4}(2p_{\nu}+k_{\nu}){\rm tr}\Big[\gamma_5\slashed{v}\frac{1}{\slashed{p}-m}\gamma^{\lambda}\frac{1}{\slashed{p}+\slashed{k}-m}\Big].
\end{eqnarray}
Here the factor $2p_{\nu}+k_{\nu}$ originates from the nonminimal vertex given by the expression $-ie\bar{\psi}\frac{\eta_2}{M} \gamma_5 \slashed{v}v^{\mu}v^{\nu}((A_{\mu}\pa_{\nu}+A_{\nu}\pa_{\mu})+(\pa_{\mu}A_{\nu}))\psi$ (the moment $p$ is for the spinor propagator, and the moment $k$ is for external gauge field). It remains to expand this expression up to the third order in external $k$ (actually, the first order in $k$ disappears, so, it remains to deal only with the third order).
Indeed, the trace in (\ref{s2}) can be calculated before of any expansion of the propagator in series in $k$:
\begin{eqnarray}
\label{s2a}
S_{2MP}&=&-4e^2\frac{\eta_2}{M}v^{\nu}v^{\mu}A_{\mu}(-k)A_{\lambda}(k)
\epsilon^{\alpha\beta\lambda\rho}
\times\nonumber\\&\times&\int\frac{d^4p}{(2\pi)^4}(2p_{\nu}+k_{\nu})v_{\alpha}p_{\beta}k_{\rho}
\frac{1}{(p^2-m^2)[(p+k)^2-m^2]}.
\end{eqnarray}
Now, this expression can be expanded into power series in external momenta, where only the third order should be taken into account (the first and second orders evidently vanish: for the first order, one evidently will have the contraction of the Levi-Civita symbol with two Lorentz-breaking vectors which immediately vanishes, and for the second order, the corresponding scalar simply does not exist).

However, to study it we can first present it as
\bea
S_{2MP}=A_{\mu}(-k)\,\Pi_{2MP}^{\mu\lambda}\,A_{\lambda}(k),
\eea
where
\begin{eqnarray}\label{Pi2MP}
\label{res}
\Pi_{2MP}^{\mu\lambda}&=&-4e^2\frac{\eta_2}{M}v^{\nu}v^{\mu}
\epsilon^{\alpha\beta\lambda\rho}k_{\rho}v_{\alpha}Q_{\nu\beta},
\end{eqnarray}
with
\begin{eqnarray}
\label{res1}
Q_{\nu\beta}&=&\int\frac{d^4p}{(2\pi)^4}(2p_{\nu}+k_{\nu})p_{\beta}
\frac{1}{(p^2-m^2)[(p+k)^2-m^2]}.
\end{eqnarray}
It is clear that, up to the second order in the external $p$ (the highest relevant order), the tensor $Q_{\nu\beta}$ must look like
\bea
Q_{\nu\beta}=Q_1\eta_{\nu\beta}+Q_2k_{\nu}k_{\beta}.
\eea
Indeed, there is no other possible tensor structures. Here $Q_1,Q_2$ are two (divergent) constants.
Substituting this structure to the contribution (\ref{res}), we find that it identically vanishes. Hence this "mixed" contribution is zero, and the only non-trivial result for the Myers-Pospelov term is given by (\ref{zzfdd}). The divergent nature of this result immediately shows that for consistency of the theory, one should have the higher-derivative CFJ-like and Myers-Pospelov term presented in the classical action from the very beginning.

Now, we can discuss the renormalization. Actually, our theory is non-renormalizable (indeed, our coupling is $\frac{\alpha}{M}$, and it has a negative mass dimension; we note that the models considered in \cite{MNP,Maluf} allowing for arising  the higher-derivative Lorentz-breaking terms are also non-renormalizable). As it is well known, the non-renormalizable theories are treated as effective ones (see a detailed discussion of the concept of effective field theories in \cite{Georgi}). They typically arise  after integrating over some fields, usually the heavy ones with the characteristic mass $M_{char}$ whose role is played in our theory by $M$, in some fundamental, renormalizable theory. As a result, the action of an effective theory represents itself as  a power series in $\frac{1}{M_{char}}$, hence the coupling constants, being proportional to different positive degrees of $\frac{1}{M_{char}}$, have negative mass dimension, and the theory turns out to be non-renormalizable. However, the linearly and quadratically divergent contributions in the effective theories are proportional to positive degrees of the cutoff scale $\mu$, and if $\mu\ll M_{char}$, these contributions turn out to be strongly suppressed being proportional to $(\frac{\mu}{M_{char}})^n$, with $n\geq 1$. Since $\frac{\mu}{M_{char}}\ll 1$, it is sufficient to restrict the expansion in $\frac{1}{M_{char}}$ by the first order. This is just the case of our theory, where $M_{char}=M$ is of the order of the Planck scale, while $\mu$ is naturally estimated to be of the order of 1 TeV, see f.e. \cite{BCP}. So, we can restrict ourselves by the contributions of the first order in $\frac{1}{M}$.

It is not difficult to show that, in the one-loop approximation,  for external $A_{\mu}$, the superficial degree of divergence looks like
\bea
\omega=4-V_1-2V_{0a}-2V_{0b}-2V_2-N_d,
\eea 
where $V_1$ is a number of vertices $\bar{\psi}\gamma_5\slashed{v} v^{\mu}v^{\nu}A_{\mu}\partial_{\nu}\psi$, $V_2$ -- of vertices $\bar{\psi}\gamma_5\slashed{v}(v\cdot A)^2\psi$,  $V_{0a}$ -- of standard vertices $\bar{\psi}\slashed{A}\psi$, $V_{0b}$ -- of vertices  $\bar{\psi}\gamma_5\slashed{v} v^{\mu}v^{\nu}(\partial_{\mu}A_{\nu})\psi$, and $N_d$ is a number of derivatives acting to external legs (except of those ones in  $V_{0b}$). We note that only $V_{0a}$ vertices are not $\frac{1}{M}$ suppressed. It is clear then that, first, the number of external $A_{\mu}$ legs cannot be less than two (hence $2V_2+V_{0a}+V_{0b}+V_1\geq 2$), and that by the gauge symmetry reasons there must be at least one derivative acting to gauge legs (to get the CFJ term) or two derivatives (to get the Maxwell or aether terms). Also, it is evident that the potentially divergent Feynman diagrams with $V_2=1,2$ will be not gauge invariant since they will yield the contributions proportional to $(v\cdot A)^2$ or $(v\cdot A)^4$, and they should vanish in some regularization. Hence in divergent diagrams one should have $V_2=0$. Then, the diagram with $V_{0a}=2$ has been studied above (\ref{zzfdd}), and the contribution with $V_{0a}=1$ and $V_{0b}+V_1=1$ is just that one given by (\ref{s2a}), and its contribution is zero. In principle we can also have divergences in contributions to the two-point function formed by only $V_1$ and $V_{0b}$ vertices, however, they are strongly suppressed being proportional to $\frac{1}{M^2}$. We conclude our discussion with the statement that up to the order $M^{-1}$, our results are exact, and the only nontrivial  divergent contribution is given by (\ref{zzfdd}).

Therefore we find that the higher-derivative action given by the sum of (\ref{free}) and (\ref{fermi}) is consistent in the one-loop order. We note that, as usual, if we suggest the gauge field to be a purely external one, the one-loop result is exact. 

Now, let us introduce finite temperature. To do it, we apply the Matsubara formalism, i.e., in the integrals over momenta above, (\ref{zaszasz}) and (\ref{Pi2MP}), we suggest the zero component of the internal momentum to be discrete, $p_0=2\pi T(l+\frac{1}{2})$, with $l$ being an integer, and, afterwards, we integrate over spatial components of the internal momentum and sum over $l$. As a result, at the finite temperature, our self-energy tensor, given by $\Pi_{MP}^{\mu \nu}=\Pi_{1MP}^{\mu \nu}+\Pi_{2MP}^{\mu \nu}$, turns out to look like
\begin{eqnarray}\label{kmT}
\Pi_{MP}^{\mu \nu} &=& A(\xi) \left[k^{2}v^{2} - 2 (v \cdot k)^{2}\right] \epsilon^{\mu\nu\lambda\rho}v_{\lambda} k_{\rho} +B(\xi)k^{2}(v\cdot t)^{2}\epsilon^{\mu\nu\lambda\rho}v_{\lambda} k_{\rho} \nonumber\\
&&-\frac12B(\xi)k^{2}v^2(t^\mu t_\alpha \epsilon^{\alpha\nu\lambda\rho}+t^\nu t_\alpha \epsilon^{\mu\alpha\lambda\rho})v_{\lambda} k_{\rho} \nonumber\\
&&-2B(\xi)(k\cdot v)(k\cdot t)(v\cdot t)\epsilon^{\mu\nu\lambda\rho}v_{\lambda} k_{\rho}\nonumber\\
&&+B(\xi)(v\cdot k)^{2}(t^\mu t_\alpha \epsilon^{\alpha\nu\lambda\rho}+t^\nu t_\alpha \epsilon^{\mu\alpha\lambda\rho})v_{\lambda} k_{\rho} \nonumber\\
&&-B(\xi)(k\cdot v)(k\cdot t)(v^\mu t_\alpha \epsilon^{\alpha\nu\lambda\rho}+v^\nu t_\alpha \epsilon^{\mu\alpha\lambda\rho})v_{\lambda} k_{\rho} \nonumber\\
&&+2B(\xi)(v\cdot k)(v\cdot t)(k^\mu t_\alpha \epsilon^{\alpha\nu\lambda\rho}+k^\nu t_\alpha \epsilon^{\mu\alpha\lambda\rho})v_{\lambda} k_{\rho} \nonumber\\
&&-\frac12B(\xi)k^2v^2(k\cdot t)\epsilon^{\mu\nu\lambda\rho}v_{\lambda} t_{\rho}\nonumber\\
&&+2B(\xi)k^2(k\cdot v)(v\cdot t)\epsilon^{\mu\nu\lambda\rho}v_{\lambda} t_{\rho}\nonumber\\
&&+C(\xi)k^{2}(v\cdot t)^2(t^\mu t_\alpha \epsilon^{\alpha\nu\lambda\rho}+t^\nu t_\alpha \epsilon^{\mu\alpha\lambda\rho})v_{\lambda} k_{\rho} \nonumber\\
&&-2C(\xi)(k\cdot t)^{2}v^2(t^\mu t_\alpha \epsilon^{\alpha\nu\lambda\rho}+t^\nu t_\alpha \epsilon^{\mu\alpha\lambda\rho})v_{\lambda} k_{\rho} \nonumber\\ 
&&-2C(\xi)(k\cdot t)^3v^2\epsilon^{\mu\nu\lambda\rho}v_{\lambda} t_{\rho} \nonumber\\ 
&&+C(\xi)k^2(k\cdot t)(v\cdot t)^2\epsilon^{\mu\nu\lambda\rho}v_{\lambda} t_{\rho} \nonumber\\ 
&&+D(\xi)(k\cdot t)^{2}(v\cdot t)^2(t^\mu t_\alpha \epsilon^{\alpha\nu\lambda\rho}+t^\nu t_\alpha \epsilon^{\mu\alpha\lambda\rho})v_{\lambda} k_{\rho} \nonumber\\
&&+D(\xi)(k\cdot t)^3(v\cdot t)^2\epsilon^{\mu\nu\lambda\rho}v_{\lambda} t_{\rho} 
+ {\cal O}\left(\frac{k^2}{m^2}\right),
\end{eqnarray}
where
\begin{eqnarray}
A(\xi) &=& -\frac{1}{12\pi^2\epsilon'} \frac{\eta_2}{M}-\frac{1}{12\pi^2} \frac{\eta_2}{M}\int_{|\xi|}^\infty dz\frac{(\tanh(\pi  z) -1)}{\sqrt{(z-\xi ) (\xi +z)}}, \\
B(\xi) &=& -\frac{1}{6}\frac{\eta_2}{M}\int_{|\xi|}^\infty dz\sqrt{(z-\xi ) (\xi +z)} \tanh(\pi  z) \text{sech}^2(\pi  z), \\
C(\xi) &=& \frac{1}{12}\frac{\eta_2}{M}\int_{|\xi|}^\infty dz\frac{\left(\xi ^2-2 z^2\right)\tanh(\pi  z)\text{sech}^2(\pi  z)}{\sqrt{(z-\xi ) (\xi +z)}}, \\
D(\xi) &=& \frac{1}{3}\frac{\eta_2}{M}\int_{|\xi|}^\infty dz\frac{ \tanh(\pi  z)\text{sech}^4(\pi  z)}{\sqrt{(z-\xi ) (\xi +z)}}\left(-5 \pi ^2 \xi ^4+\xi ^2+\left(5 \pi ^2 \xi ^2-2\right) z^2 \right. \nonumber\\
&&\left.+\left(\pi ^2 \xi ^4+\xi ^2-\left(\pi ^2 \xi ^2+2\right) z^2\right)\cosh(2 \pi  z)\right),
\end{eqnarray}
with $\xi=\frac{m}{2\pi T}$. Above, we have split the internal momentum as $p^\mu=\vec p^\mu + p_0 t^\mu$, with $\vec p^\mu=(0,\vec p)$ and $t^{\mu}=(1,0,0,0)$ being a constant vector along the time axis. We note that these functions of the temperature emerged as well in \cite{MNPT} where the finite-temperature extension of results obtained in \cite{MNP} for another Lorentz-breaking extension of the QED, involving the magnetic coupling and the coupling of $\psi$ to the constant axial vector $b_{\mu}$, was carried out. It was shown there that in the high temperature limit all these functions vanish. The result (\ref{kmT}) is clearly gauge invariant.

Besides of the two-point function of the gauge field, it is important also to consider the two-point function of the spinor field. 

Let us calculate this two-point function of the spinor in the first order in $\frac{1}{M}$.
We start with the action given by the sum of Lorentz-breaking classical actions (\ref{quad}) and (\ref{fermi}) allowing to obtain the propagators of gauge and spinor fields respectively.
The two-point function of the spinor field is generated by two contributions with external spinor legs: the first one involves two triple vertices, and the second one involves one quartic vertex.
To do the calculation, we proceed in the same manner as with the two-point function of the gauge field, that is, we note that the result can be represented in the form of the expansion in $\frac{1}{M}$, and will find the first order in this expansion, just as we have done above. The Lorentz-breaking insertions into the Feynman diagrams below are denoted by $\times$ symbol.

\begin{figure}[htbp]
\centering
\includegraphics{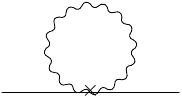}
\caption{\label{Fig3} The contributions to the two-point function of $\psi$ with the quartic vertex.}
\end{figure}

We see that since the quartic vertex is proportional to $\frac{1}{M}$, we can keep in the propagator of the gauge field only the zero-order terms in $\frac{1}{M}$, so, we have $<A^{\mu}(-k)A^{\nu}(k)>=\frac{i\eta^{\mu\nu}}{k^2}$. Hence, the contribution of the diagram with quartic vertex given by Fig.~\ref{Fig3} is proportional to $\int\frac{d^4k}{(2\pi)^4}\frac{1}{k^2}=0$. As a result, we are left with triple vertices only. They look like 
\begin{eqnarray}
{\cal L}_{triple}&=&-e\bar{\psi}\Big(-\slashed{A}+i\frac{\eta_2}{M} \gamma_5 \slashed{v}v^{\mu}v^{\nu}(A_{\mu}\pa_{\nu}+A_{\nu}\pa_{\mu}+(\pa_{\mu}A_{\nu}))
\Big)\psi.
\end{eqnarray}

The explicit form of the vertices, in the momentum space, is
\bea
V_0(l_1,l_2,l_3)&=&e\bar{\psi}(l_1)\gamma^{\kappa}\psi(l_2)A_{\kappa}(l_3)(2\pi)^4\delta(l_1+l_2+l_3),\nonumber\\
V_1(k_1,k_2,k_3)&=&-\frac{e\eta_2}{M}v^{\mu}v^{\nu}\bar{\psi}(k_1)\gamma_5\vs\psi(k_2)A_{\lambda}(k_3)[\delta^{\lambda}_{\mu}k_{2\nu}-\delta^{\lambda}_{\nu}k_{1\mu}](2\pi)^4\delta(k_1+k_2+k_3).
\eea

\begin{figure}[htbp]
\centering
\includegraphics{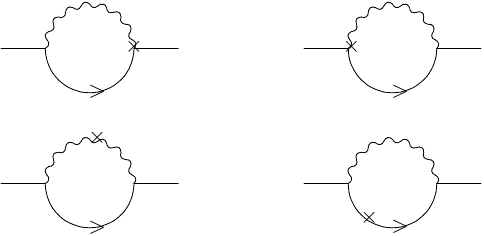}
\caption{\label{Fig4} The contributions to the two-point function of $\psi$ with triple vertices.}
\end{figure}

So we have the graphs given by Fig.~\ref{Fig4}. In two upper graphs of Fig.~\ref{Fig4} we consider usual propagators and modified vertices (i.e. one usual vertex $V_0$ and one new vertex $V_1$). The result is, respectively:
\bea
T_1(k)&=&-\frac{e^2\eta_2}{M}v^{\mu}v^{\nu}\int\frac{d^4p}{(2\pi)^4}\bar{\psi}(-k)\gamma^{\kappa}(\ps+m)\gamma_5\vs\psi(k)\eta_{\lambda\kappa}[\delta^{\lambda}_{\mu}k_{\nu}-\delta^{\lambda}_{\nu}p_{\mu}]\frac{1}{(p^2-m^2)(k+p)^2},\nonumber\\
T_2(k)&=&-\frac{e^2\eta_2}{M}v^{\mu}v^{\nu}\int\frac{d^4p}{(2\pi)^4}\bar{\psi}(-k)
\gamma_5\vs(\ps+m)\gamma^{\kappa}\psi(k)\eta_{\lambda\kappa}[\delta^{\lambda}_{\nu}k_{\mu}-\delta^{\lambda}_{\mu}p_{\nu}]\frac{1}{(p^2-m^2)(k+p)^2}.
\eea

In two lower graphs of Fig.~\ref{Fig4} we have insertions into the propagators. 
The graph with an insertion into the gauge propagator, with $\Sigma(k)=-c_2u^2k^2+c_1(u\cdot k)^2$, is
\bea
T_3(k)&=&\frac{4e^2\eta_2}{M}\int\frac{d^4p}{(2\pi)^4}\bar{\psi}(-k)\gamma^{\mu}(\ps+m)\gamma^{\nu}\psi(k)\frac{\eta_{\nu\alpha}\eta_{\mu\beta}}{(k+p)^4}\epsilon^{\alpha\beta\rho\sigma}u_{\rho}(p_{\sigma}+k_{\sigma})\frac{\Sigma(k+p)}{(p^2-m^2)}.
\eea
And the graph with an insertion into the spinor propagator is
\bea
T_4(k)&=&\frac{e^2\eta_2}{M}\int\frac{d^4p}{(2\pi)^4}\bar{\psi}(-k)\gamma^{\mu}(\ps+m)\gamma_5\vs(v\cdot p)^2(\ps+m)\gamma^{\nu}\psi(k)\frac{\eta_{\mu\nu}}{(k+p)^2}
\frac{1}{(p^2-m^2)^2}.
\eea
Within the calculation, we expand these contributions up to the second order in the external $k_{\mu}$. It remains to simplify these expressions. Again, we use the dimensional regularization with $\epsilon=4-D$. We take into account only terms up to the second order in external momenta, because the higher orders do not contribute to one-loop divergences.

It is instructive here to give some intermediate steps of the calculation.
First of all, the structures of $T_1$ and $T_2$ are rather similar, so that can be summed, and $T_4$ can be simplified with use of the identities $\gamma^{\mu}\gamma^{\nu}\gamma_{\mu}=-2\gamma^{\nu}$ and $\gamma^{\mu}[\gamma^{\alpha},\gamma^{\beta}]\gamma_{\mu}=0$. Then, we have
\bea
T_1(k)+T_2(k)&=&-\frac{2e^2\eta_2}{M}\int\frac{d^4p}{(2\pi)^4}\bar{\psi}(-k)\vs\ps\vs\gamma_5\psi(k)
\frac{v\cdot(p-k)}{(p^2-m^2)(k+p)^2},\nonumber\\
T_4(k)&=&\frac{2e^2\eta_2}{M}\int\frac{d^4p}{(2\pi)^4}\bar{\psi}(-k)
[2(v\cdot p)\ps-\vs(p^2+m^2)]\gamma_5
\psi(k)\frac{(v\cdot p)^2}{(k+p)^2}
\frac{1}{(p^2-m^2)^2}.
\eea
Integrating over momenta, we arrive at the following pole parts of these contributions up to the second order in the external $k_\mu$, either in the massive case, where one can use expansion in $\frac{k^2}{m^2}$,
\bea
\label{t12mn}
T_1(k)+T_2(k)&=&\frac{i e^2 \eta _2}{48 \pi ^2 \epsilon' M}\bar{\psi}(-k)\left[\left(v^2 \left(k^2-3 m^2\right)-20 (k\cdot v)^2\right) \slashed{v}+10 v^2 (k\cdot v) \slashed{k}\right]\gamma_5\psi(k) \nonumber\\
&&-\frac{i e^2 \eta _2}{576 \pi ^2 M} \bar{\psi}(-k)\left(-5 k^2 v^2 \slashed{v}-98 v^2 (k\cdot v) \slashed{k}+196 (k\cdot v)^2 \slashed{v}\right)\gamma_5\psi(k) \nonumber\\
&&-\frac{3 i e^2 \eta _2}{64 \pi ^2 M}\bar\psi(-k)m^2 v^2 \slashed{v}\gamma_5\psi(k) + {\cal O}\left(\frac{k^2}{m^2}\right),
\eea
or in the massless limit, where one uses expansion in $\frac{m^2}{k^2}$,
\bea
\label{t12m0}
T_1(k)+T_2(k)&=&\frac{i e^2 \eta _2}{48 \pi ^2 \epsilon'' M}\bar{\psi}(-k)\left[\left(v^2 k^2-20 (k\cdot v)^2\right) \slashed{v}\gamma ^5+10 v^2 (k\cdot v) \slashed{k}\gamma_5\right]\psi(k) \nonumber\\
&&+\frac{i e^2 \eta _2}{288 \pi ^2 k^6 M}\bar{\psi}(-k)\left[62 k^6 v^2 (k\cdot v) \slashed{k}\gamma ^5+\slashed{v}\gamma_5 \left(8k^8 v^2-124 k^6 (k\cdot v)^2\right)\right]\psi(k) \nonumber\\
&&+ {\cal O}\left(\frac{m^2}{k^2}\right).
\eea
For $T_4$, in massive and massless cases, respectively, we have
\bea
\label{t4mn}
T_4(k)&=&-\frac{i e^2 \eta _2}{96 \pi ^2 \epsilon' M}\bar\psi(-k)\left[v^2 \left(\left(k^2-6 m^2\right) \slashed{v}-6 \slashed{k} (k\cdot v)\right)+2 (k\cdot v)^2 \slashed{v}\right]\gamma_5\psi(k) \nonumber\\
&&-\frac{i e^2 \eta _2}{2304 \pi ^2 M}\bar\psi(-k)\left(k^2 v^2 \slashed{v}-78 v^2 \slashed{k} (k\cdot v)+98 (k\cdot v)^2 \slashed{v}\right)\gamma_5\psi(k) \nonumber\\
&&+\frac{7 i e^2 \eta _2}{192 \pi ^2 M}\bar\psi(-k)m^2 v^2 \slashed{v}\gamma_5\psi(k) + {\cal O}\left(\frac{k^2}{m^2}\right)
\eea
or
\bea
\label{t4m0}
T_4(k)&=&-\frac{i e^2 \eta _2}{96 \pi ^2 \epsilon'' M}\bar\psi(-k)\left[v^2 \left( k^2 \slashed{v}-6 \slashed{k} (k\cdot v)\right)+2 (k\cdot v)^2 \slashed{v}\right]\gamma_5\psi(k) \nonumber\\
&&-\frac{i e^2 \eta _2}{1152 \pi ^2 k^{10} M}\bar\psi(-k)\left[k^2 v^2 \left(16 k^{10} \slashed{v}-60 k^8 \slashed{k} (k\cdot v)\right)+2 (k\cdot v)^2 \left(22 k^{10} \slashed{v}-\right.\right.\nonumber\\ &-&\left.\left.36 k^8 \slashed{k} (k\cdot v)\right)\right]\gamma_5\psi(k) + {\cal O}\left(\frac{m^2}{k^2}\right).
\eea

The $T_3$ has a structure different from $T_1,T_2,T_4$, being proportional to the Levi-Civita symbol.  After the integrations over momenta we find for massive and massless cases, respectively,
\bea
T_3(k)&=&-\frac{i e^2 \eta _2}{96 \pi ^2 \epsilon' M}u^\kappa  \epsilon _{\kappa \lambda \mu  \nu}\bar\psi(-k)\left[-\gamma ^{\mu }\gamma ^{\nu }\gamma ^{\lambda } \left(u^2 \left(c_1 \left(2 m^2-k^2\right)+4 c_2 \left(k^2-3 m^2\right)\right)+2 c_1 (k\cdot u)^2\right) \right. \nonumber\\
&&\left.+\left(2 u^2  \left(\left(c_1-4 c_2\right) \slashed{k}+2 \left(c_1-6 c_2\right) m\right)+4 c_1 (k\cdot u)  \slashed{u}\right)k^\lambda\gamma ^{\mu}\gamma ^{\nu}\right]\psi(k) \nonumber\\
&&-\frac{i e^2 \eta _2}{2304 \pi ^2 M}\bar\psi(-k)\gamma ^{\mu }\gamma ^{\nu }\gamma ^{\lambda }\psi(k) \left(-40 c_2 k^2 u^2+13 c_1 k^2 u^2-14 c_1 (k\cdot u)^2\right) u^\kappa \epsilon _{\kappa \lambda  \mu  \nu} \nonumber\\
&&-\frac{i e^2 \eta _2}{1152 \pi ^2 M}\left(13 c_1-40 c_2\right)u^2 u^\kappa k^\lambda \epsilon _{\kappa \lambda \mu  \nu} \bar\psi(-k)\slashed{k}\gamma ^{\mu }\gamma ^{\nu }\psi(k) \nonumber\\
&&+\frac{7 i c_1 e^2 \eta _2}{576 \pi ^2 M} (k\cdot u) u^\kappa k^\lambda \epsilon _{\kappa \lambda \mu  \nu} \bar\psi(-k)\slashed{u}\gamma ^{\mu }\gamma ^{\nu }\psi(k) \nonumber\\
&&-\frac{i e^2 \eta _2}{288 \pi ^2 M}\left(5 c_1-18 c_2\right)m u^2 \bar\psi(-k)\gamma ^{\mu }\gamma ^{\nu }\psi(k) u^\kappa k^\lambda \epsilon _{\kappa \lambda \mu  \nu} \nonumber\\
&&+\frac{i e^2 \eta _2}{576 \pi ^2 M}\left(11 c_1-54 c_2\right)m^2u^2\bar\psi(-k)\gamma ^{\mu }\gamma ^{\nu }\gamma ^{\lambda }\psi(k) u^\kappa \epsilon _{\kappa \lambda  \mu  \nu} +{\cal O}\left(\frac{k^2}{m^2}\right)
\eea
or
\bea
T_3(k)&=&-\frac{i e^2 \eta _2}{96 \pi ^2 \epsilon' M}u^\kappa  \epsilon _{\kappa \lambda \mu  \nu}\bar\psi(-k)\left[-\gamma ^{\mu }\gamma ^{\nu }\gamma ^{\lambda } \left(u^2 \left(-c_1+4 c_2\right)k^2+2 c_1 (k\cdot u)^2\right) \right. \nonumber\\
&&\left.+k^\lambda \left(2 u^2 \left(c_1-4 c_2\right) \slashed{k}+4 c_1 (k\cdot u) \slashed{u}\right)\gamma ^{\mu }\gamma ^{\nu }\right]\psi(k) \nonumber\\
&&-\frac{i e^2 \eta _2}{576 \pi ^2 k^{10} M}u^\kappa k^\lambda \epsilon _{\kappa \lambda \mu  \nu} \left(k^2 u^2 \left(10 c_1 k^8 -40 c_2 k^8 \right)+12 c_1 k^8 (k\cdot u)^2\right) \bar\psi(-k)\slashed{k}\gamma ^{\mu }\gamma ^{\nu }\psi(k) \nonumber\\
&&+\frac{i e^2 \eta _2}{1152 \pi ^2 k^8 M}\bar\psi(-k)\gamma ^{\mu }\gamma ^{\nu }\gamma ^{\lambda }\psi(k)u^\kappa \epsilon _{\kappa \lambda  \mu  \nu} \left(k^2 u^2 \left(-16 c_1 k^8 +64 c_2 k^8 \right)+20 c_1 k^8 (k\cdot u)^2\right) \nonumber\\
&&+\frac{5i c_1 e^2 \eta _2}{144 \pi ^2 M}(k\cdot u) u^\kappa k^\lambda \epsilon _{\kappa \lambda \mu  \nu} \bar\psi(-k)\slashed{u}\gamma ^{\mu }\gamma ^{\nu }\psi(k) +{\cal O}\left(\frac{m^2}{k^2}\right).
\eea
However, the form of $T_3$ can be reduced to that one similar to that of $T_1,T_2,T_4$, with the use of the identities:
\bea
\label{ids}
\sigma^{\mu\nu}\gamma_5&=&\frac{i}{2}\epsilon^{\mu\nu\alpha\beta}\sigma_{\alpha\beta},\quad\, \epsilon_{\kappa\lambda\mu\nu}\gamma^{\mu}\gamma^{\nu}=-2\sigma_{\kappa\lambda}\gamma_5,\nonumber\\
\epsilon_{\kappa\lambda\mu\nu}\gamma^{\lambda}\gamma^{\mu}\gamma^{\nu}&=&-6i\gamma_5\gamma_{\kappa},
\eea
which implies, at $m\neq 0$,
\bea
\label{t3mn}
T_3(k)&=&-\frac{ e^2 \eta _2}{96 \pi ^2 \epsilon' M} \bar\psi(-k)\left[
6\gamma_5\us\left(u^2 \left(c_1 \left(2 m^2-k^2\right)+4 c_2 \left(k^2-3 m^2\right)\right)+2 c_1 (k\cdot u)^2\right) \right. \nonumber\\
&&\left.-2i[2 u^2  \left(\left(c_1-4 c_2\right) \slashed{k}+2 \left(c_1-6 c_2\right) m \right)+4 c_1 (k\cdot u)  \slashed{u}]
u^{\kappa}k^{\lambda}\sigma_{\kappa\lambda}\gamma_5\right]\psi(k) \nonumber\\
&&-\frac{e^2 \eta _2}{384 \pi ^2 M}\bar\psi(-k)\gamma_5\us\psi(k) \left((13 c_1-40c_2) k^2 u^2-14 c_1 (k\cdot u)^2\right)  \nonumber\\
&&+i\frac{e^2 \eta _2}{576 \pi ^2 M}\left(13 c_1-40 c_2\right)u^2 u^\kappa k^\lambda  \bar\psi(-k)\slashed{k}\sigma_{\kappa\lambda}\gamma_5\psi(k) \nonumber\\
&&-\frac{7i c_1 e^2 \eta _2}{288 \pi ^2 M} (k\cdot u) u^\kappa k^\lambda \bar\psi(-k)\slashed{u}\sigma_{\kappa \lambda}\gamma_5 \psi(k) \nonumber\\
&&+\frac{ie^2 \eta _2}{144 \pi ^2 M}\left(5 c_1-18 c_2\right)m u^2 \bar\psi(-k)\sigma_{\kappa\lambda}\gamma_5\psi(k) u^\kappa k^\lambda \nonumber\\
&&+\frac{e^2 \eta _2}{96 \pi ^2 M}\left(11 c_1-54 c_2\right)m^2u^2\bar\psi(-k)\gamma_5\us\psi(k) +{\cal O}\left(\frac{k^2}{m^2}\right)
\eea
and, at $m\to 0$,
\bea
\label{t3m0}
T_3(k)&=&-\frac{ e^2 \eta _2}{96 \pi ^2 \epsilon' M}\bar\psi(-k)\left[6\gamma_5\us\left(u^2 \left(-c_1+4 c_2\right)k^2+2 c_1 (k\cdot u)^2\right) \right. \nonumber\\
&&\left.-4i  u^{\kappa}k^\lambda (u^2\left((c_1-4 c_2) \slashed{k}\right) +4 c_1 (k\cdot u) \slashed{u})\sigma_{\kappa\lambda}\gamma_5\right]\psi(k) \nonumber\\
&&+\frac{ie^2 \eta _2}{288 \pi ^2 k^{10} M}u^\kappa k^\lambda \left(k^2 u^2 \left(10 c_1 k^8 -40 c_2 k^8 \right)+12 c_1 k^8 (k\cdot u)^2\right) \bar\psi(-k)\slashed{k}\sigma_{\kappa\lambda}\gamma_5\psi(k) \nonumber\\
&&-\frac{e^2 \eta _2}{192 \pi ^2 k^8 M}\bar\psi(-k))\us\gamma_5\psi(k) \left(k^2 u^2 \left(-16 c_1 k^8 +64 c_2 k^8 \right)+20 c_1 k^8 (k\cdot u)^2\right) \nonumber\\
&&-\frac{5ic_1 e^2 \eta _2}{72 \pi ^2 M}(k\cdot u) u^\kappa k^\lambda \bar\psi(-k)\slashed{u}\sigma_{\kappa\lambda}\gamma_5\psi(k) +{\cal O}\left(\frac{m^2}{k^2}\right).
\eea
More simplifications are possible in terms involving $\sigma_{\nu\lambda}$ matrices, due to symmetrization by the rules like $\ks\ks=k^2$, $\vs\vs=v^2$, and then, $u^{\kappa}k^{\lambda}\ks\sigma_{\kappa\lambda}=i[(u\cdot k)\ks-\us k^2]$, and $u^{\kappa}k^{\lambda}\us\sigma_{\kappa\lambda}=i[u^2\ks-\us(u\cdot k)]$. 
Thus, we find
\bea
\label{t3mna}
T_3(k)&=&-\frac{ e^2 \eta _2}{96 \pi ^2 \epsilon' M} \bar\psi(-k)\left[
6\us\left(u^2 \left(c_1 \left(2 m^2-k^2\right)+4 c_2 \left(k^2-3 m^2\right)\right)+2 c_1 (k\cdot u)^2\right) \right. \nonumber\\
&&\left.+4 u^2  \left(\left(c_1-4 c_2\right) [(u\cdot k)\ks-\us k^2]+2i \left(c_1-6 c_2\right) m u^{\kappa}k^{\lambda}\sigma_{\kappa\lambda}\right)+\right.\nonumber\\&&\left.
+4 c_1 (k\cdot u) [u^2\ks-\us(u\cdot k)]
\right]\gamma_5\psi(k) \nonumber\\
&&-\frac{e^2 \eta _2}{384 \pi ^2 M}\bar\psi(-k)\gamma_5\us\psi(k) \left((13 c_1-40c_2) k^2 u^2-14 c_1 (k\cdot u)^2\right)  \nonumber\\
&&-\frac{e^2 \eta _2}{576 \pi ^2 M}\left(13 c_1-40 c_2\right)u^2   \bar\psi(-k)[(u\cdot k)\ks-\us k^2]\gamma_5\psi(k) \nonumber\\
&&+\frac{7c_1 e^2 \eta _2}{288 \pi ^2 M} (k\cdot u)  \bar\psi(-k)[u^2\ks-\us(u\cdot k)]\gamma_5 \psi(k) \nonumber\\
&&+\frac{ie^2 \eta _2}{144 \pi ^2 M}\left(5 c_1-18 c_2\right)m u^2 \bar\psi(-k)\sigma_{\kappa\lambda}\gamma_5\psi(k) u^\kappa k^\lambda \nonumber\\
&&+\frac{e^2 \eta _2}{96 \pi ^2 M}\left(11 c_1-54 c_2\right)m^2u^2\bar\psi(-k)\gamma_5\us\psi(k) +{\cal O}\left(\frac{k^2}{m^2}\right)
\eea
and, at $m\to 0$,
\bea
\label{t3m0a}
T_3(k)&=&-\frac{ e^2 \eta _2}{96 \pi ^2 \epsilon' M}\bar\psi(-k)\left[6\us\left(u^2 \left(-c_1+4 c_2 \right)k^2+2 c_1 (k\cdot u)^2\right) \right. \nonumber\\
&&\left.+4u^2  (c_1-4 c_2)[(u\cdot k)\ks-\us k^2]+4 c_1 (k\cdot u) [u^2\ks-\us(u\cdot k)]\right]
\gamma_5\psi(k) \nonumber\\
&&-\frac{e^2 \eta _2}{288 \pi ^2 k^2 M} \left(k^2 u^2 \left(10 c_1-40 c_2 \right)+12 c_1  (k\cdot u)^2\right) \bar\psi(-k)[(u\cdot k)\ks-\us k^2]\gamma_5\psi(k) \nonumber\\
&&-\frac{e^2 \eta _2}{192 \pi ^2 M}\bar\psi(-k))\us\gamma_5\psi(k) \left(k^2 u^2 \left(-16 c_1 +64 c_2  \right)+20 c_1  (k\cdot u)^2\right) \nonumber\\
&&+\frac{5c_1 e^2 \eta _2}{72 \pi ^2 M}(k\cdot u)  \bar\psi(-k)[u^2\ks-\us(u\cdot k)]\gamma_5\psi(k) +{\cal O}\left(\frac{m^2}{k^2}\right).
\eea

Taking all together, we find that, to achieve multiplicative renormalizability, in $m\neq 0$ case, the total free Lorentz-breaking Lagrangian of the spinor, corresponding to pole parts of $T_1,T_2,T_3,T_4$ together plus the classical action, must be
\bea
\label{totall}
{\cal L}_{total}&=&\bar{\psi}\left( 
i\ds-m\right)\psi+\nonumber\\
&+&\frac{1}{M}\bar{\psi}\left(C_1v^2\vs\Box+C_2(v\cdot\pa)^2\vs+C_3 v^2(v\cdot\pa)\ds+
C_4 
m^2v^2\vs+C_5m v^2\sigma_{\lambda\rho}v^{\lambda}\pa^{\rho}
\right)\gamma_5\psi +\nonumber\\ &+&
(C_i\to C_i^{\prime},v^{\mu}\to u^{\mu}),
\eea
where $C_1\ldots C_5$, $C'_1\ldots C'_5$ are dimensionless constants, and each term of the given dependence in $v^{\mu}$, has its analogue where $v^{\mu}$ is replaced by the $u^{\mu}$.  In our case, the last term proportional to $C'_5$ emerges only with $u^{\mu}$ vector, arising from $T_3$, with there is no term proportional to $C_5$. We note that, as it frequently occurs, the Lorentz-breaking vectors are light-like, some terms in quantum corrections simply vanish, so the structure of quantum corrections simplifies drastically  (f.e. the similar situation takes place in \cite{MNP}). Namely, if both $v^2=0$ and $u^2=0$, this Lagrangian exactly matches the kinetic part of the Lagrangian (\ref{fermi}) which we used as a starting point. We note that, from dimensional and symmetry reasons it is easy to conclude that the same quantum corrections (\ref{totall}) will emerge if, instead of (\ref{fermi}) we used the gauge extension of (\ref{totall}). Also, the new terms proportional to $\frac{m}{M}u^2\bar{\psi}\gamma_5(u\cdot \pa)\psi$ or its analogue where $u^{\mu}$ is replaced by $v^{\mu}$, can arise in these cases.

We note that in principle the explicit results of integration over momenta can be obtained as well in general case, without imposing any of these limits, however, they are extremely cumbersome.
It is interesting to observe that if the Lorentz-breaking vectors $u^{\mu},v^{\mu}$ are light-like, the zero and first orders in external momenta in these contributions vanish.

The whole contribution to the two-point function of the spinor is given by the sum of $T_1$, $T_2$, $T_3$ and $T_4$: in the massive case, one finds a sum of (\ref{t12mn},\ref{t3mna},\ref{t4mn}), and in the massless case, one looks the sum of (\ref{t12m0},\ref{t3m0a},\ref{t4m0}). We close the section with the conclusion that we found the two-point functions both in gauge and matter sectors of our extension of the QED. 

\section{Unitarity aspects in the extended QED}\label{IV}

It is well known that the presence of higher time derivatives in quantum field theory
can lead to an indefinite metric in Hilbert space. The sector with negative metric of the theory produces 
 negative norm states or ghosts which introduce several conceptual issues in connection with 
  the conservation of probability
or unitarity. However, in the subclass of higher time derivative theories called Lee-Wick theories, where 
the additional degrees of freedom arise 
 in complex conjugate poles, 
perturbative 
unitarity has been well established \cite{Lee-Wick}.
The idea is that since the structure of poles 
determines the discontinuities in phase space, under some assumptions both contributions of 
complex conjugate modes cancel each other order by order in the
 perturbative series \cite{Cut}. 
 The issue of analyticity 
 in the complex energy plane and the resulting cutting equations 
 have been intensively studied over the past years (for the general 
 discussion of ghost states see f.e. \cite{ghosts}). 
 The Lee-Wick prescription of removing
 the negative metric particles from the asymptotic space 
has been shown to be an efficient tool in providing a unitary theory together with the expected 
convergence property.

In general, to study unitarity in higher time derivatives theories one is confronted with the problem
of analyticity of amplitude diagrams. The direct application of the $i\epsilon$ prescription in the propagators
seems to fail to preserve unitarity in many cases, therefore it is necessary
 to analyze the configuration of poles case by case.
Moreover, the presence of Lorentz symmetry breaking makes the
 study of analyticity of integrals to be more involved.
In many cases, there can be an arbitrary number of extra poles 
associated to negative-metric states, which marks a departure 
with respect to the pole structure of a Lee-Wick theory that one uses to prove unitarity.
It is also difficult to deal with the perturbative solutions which  
 can become complex under certain conditions, and the corresponding dispersion relation can be
extremely difficult to solve
 \cite{LIV_Unit}. 
 An early approach  to deal with analytic properties of phase space
 integrals  in the presence of Lorentz violation, based on 
the Euclidean space or Wick rotation, has been presented in~\cite{Maniatis-Reyes}.
Recently a new formulation for Lee-Wick theories as non-analytical Euclidean 
theories has been proposed in~\cite{Piva-Anselmi1,Piva-Anselmi}.
We follow similar lines 
 to deal with unitarity in our
 higher-time derivative Lorentz-violating model. 
The strategy we  pursue to compute the relevant contributions of discontinuities is to
consider the Euclidean theory from the beginning and perform the Wick rotation 
together with rotation of the preferred four vector and and to apply the Lee-Wick prescription 
in cut integrals~\cite{Lee-Wick}. In this way we arrive at the simplified 
integral with simplified poles. 

The processes we study
are the Bhabha scattering at tree level (we note that some earlier 
studies of Bhabha scattering in a Lorentz-breaking extension of QED were 
carried out in~\cite{Bhabha}, where, however, no higher-derivative terms were studied) 
and Compton scattering at the one-loop level. 
In the first case, we let the preferred four vector to be the most general one, allowing additional 
degrees of freedom and the negative metric to 
arise,
and in the second one, we choose a purely 
time-like preferred four vector without ghosts in the theory.
For both cases, we consider  the forward scattering of two particles with 
incoming momenta 
$p=k$ and $p'$ related as
\begin{eqnarray}
p+p'\to p+p' \,.
\end{eqnarray}
\subsection{Bhabha scattering at tree level in the ghost sector }
\begin{figure}[h]
\centering
\includegraphics[width=0.45\textwidth]{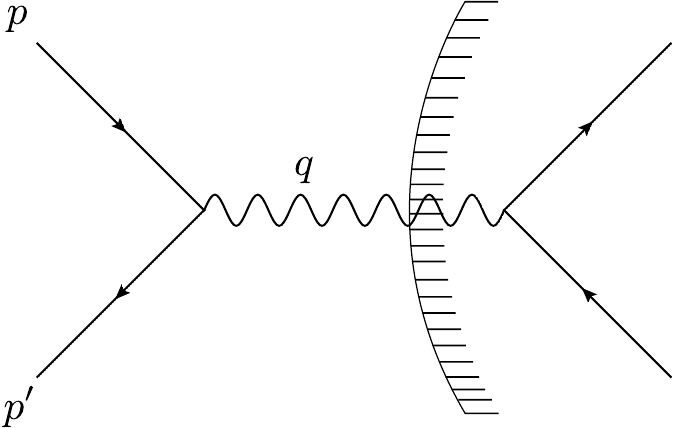}
\caption{\label{Fig1b} The Bhabha scattering diagram at tree level.}
\end{figure}

We consider a generic preferred four vector $u_{\mu}$, so that, in general, ghosts 
can arise. We also consider the Bhabha scattering
process
at tree level given by the Fig.~\ref{Fig1b}.

The amplitude in the transverse gauge is given by
\begin{eqnarray}\label{amp}
i \mathcal M=(-ie)V^{\mu} (p,p')
\times \sum _{\lambda} \frac{   -i   P^{(\lambda)}_{\mu \nu}({q})    }{
 \Lambda^{(\lambda)}(q) } \bigg\rvert_{q= p-p'} \times (-ie)V^{*\nu}  (p,p')\,,
\end{eqnarray}
where $\Lambda^{(\lambda)}(q)= q^2+\lambda a\sqrt{D(q)}$, with $D(q)$ and $a$ given by~(\ref{D}) and (\ref{Da}) respectively and 
\begin{eqnarray}
V^{ \mu}(p,p')=  \bar u(p) \gamma^{\mu}v(p^{\prime})\,,
\nonumber \qquad
V^{*\nu}(p,p')= \bar v(p^{\prime}) \gamma^{\nu}u(p)\,.
\end{eqnarray}
The standard way to compute the imaginary part of the amplitude in Eq.~\eqref{amp}, 
is to fix the four-vector $u^{\mu}$, 
solve the dispersion relation,
and, afterwards, analyze discontinuities of $\mathcal M(s)$ which is an analytic function of the complex variable $s$.
However, in our 
model with modified photons, the dispersion relation is a very complicated expression and the solutions can be 
difficult to find. So,
we introduce a novel method to deal with unitarity.
 
The strategy is to start with a theory in Minkowski space, which is defined 
as the one obtained from the Wick rotation in the Euclidean theory, 
perhaps non-analytically, as in the Ref.~\cite{Piva-Anselmi1,Piva-Anselmi}. 
This starting point ensures a well defined 
Wick rotation to the Euclidean space warranted by the position 
of positive and negative poles in the fourth and second
quadrants of the energy plane respectively. Hence, we perform the Wick rotation, changing external momenta $s_{4}=is_0$,
so that the dispersion relation 
 decouples into usual and ghost solutions. This last step 
 simplifies the calculation considerably.
The rotated energy integral 
will still depend on the $i\epsilon$ prescription which allows us to compute the discontinuity. 
Finally, we perform the polarization sum and conveniently evaluate with
 the delta function in some parts of the  integral. Only at the final step we 
 remove the $i\epsilon$ prescription performing the limit $\epsilon\to 0$.

From Eq.~(\ref{amp}), we can write 
\begin{eqnarray}\label{firstsum}
 \mathcal M(s)=  e^2V^{\mu} V^{*\nu}  \times \int \frac{d^4 q }{(2\pi)^4}
\left [\sum _{\lambda}\frac{  e_{\mu \nu}+ i \lambda \epsilon_{\mu \nu }  }{
 2  \Lambda^{(\lambda)}(q)    }  \right] _{q^2\to q^2+i\epsilon} \delta^{(4)}(q-s)\,,
\end{eqnarray}
where we have defined $s=p-p'$, included the $i\epsilon$ prescription and used $P^{(\lambda)}_{\mu \nu}
=\frac{1}{2}(e_{\mu \nu }+i \lambda \epsilon  _{\mu \nu})$. Using the expressions 
\begin{eqnarray}
\frac{1}{\Lambda^{(+)}}+\frac{1}{\Lambda^{(-)}}&=& \frac{2q^2}{(q^2)^2-a^2D}\,,  \\
\frac{1}{\Lambda^{(+)}}-\frac{1}{\Lambda^{(-)}}&=&-\frac{2a\sqrt{D}}{(q^2)^2-a^2D}\,, \nonumber
\end{eqnarray}
 and \eqref{rele1}, \eqref{rele2}, we can write
 \begin{eqnarray}
 \mathcal M(s)&=&e^2V^{\mu} V^{*\nu}  \times \int \frac{d^4 q }{(2\pi)^4}
\left[\frac{1}{(q^2)^2-a^2D}   \left( -q^2 \eta_{\mu \nu}- \frac{(q^2)^2}{D}u_{\mu}  u_{\nu}  
\right. \right.  \nonumber \\     &+&  \left. \left.ai \epsilon_{\mu \alpha \beta \nu}u^{\alpha} q^{\beta}
  \right)\right]_{q^2\to q^2+i\epsilon}  \delta^{(4)}(q-s) \,,
\end{eqnarray} 
where the terms in \eqref{firstsum} proportional to $\sla{q}=\sla{p}-\sla {p}^{\prime}$ 
vanish due to the external on-shell spinors.

The Wick rotation has to be done carefully, since the direct 
analytical extension of momentum variable in the delta can lead to inconsistencies.
The best way to proceed for our integral
 is to perform the analytic extension in the original expression \eqref{amp} 
and then go back with the integral in Eq. \eqref{firstsum}. However, as an intermediate step, we will extend the delta to complex 
variables~\cite{Piva-Anselmi}.
Before doing this, however, we should mention 
that solutions in Euclidean space may differ 
from those in Lorentzian space, so the equivalence of both methods holds 
with respect to the type of solutions which eventually propagate through the cuts.
Along these steps, by performing the analytic extension with the rule $s_0=-is_{4}$ and momenta $s_E=(\vec s,s_4)$, we arrive at
\begin{eqnarray}\label{I_Ec}
 \mathcal M(s_E)&=&e^2V^{\mu} V^{*\nu}  \times \int \frac{d^4 q_E }{(2\pi)^4} \left[  \frac{1}{q_E^2  (1+
 \beta^2 \gamma q_E^2)}   \left(  \eta_{\mu \nu}+ \frac{1}{\gamma}u^E_{\mu}  u^E_{\nu}  
 +\beta i \epsilon_{\mu \alpha \beta \nu}u_E^{\alpha} q_E^{\beta}  \right)\right]  _{q^2_E\to q^2_E-i\epsilon}   \nonumber  \\ &\times&
     \delta^{(3)}(\vec q-\vec p-\vec p') 
    \delta( q_{4}- s_4 ) \,,
\end{eqnarray}
with
\begin{eqnarray}
D_E  &=&-\gamma q_E^2\,,  \nonumber  \\  a_E&=&\beta q_E^2\,,
   \nonumber \\  \nonumber
  \gamma&=&u_E^2\sin^2\theta\,,    \\     \beta&=&\frac{4u_E^2(c_1\cos^2\theta-c_2)}{M} \,.
\end{eqnarray}
where $\theta$ is the angle between the two Euclidean four-vectors $u_E$ and $q_E$.
Now, in terms of $\varepsilon^{(\lambda)}_{E \nu }$ which is a function of $q_{4}$ we can 
write Eq. \eqref{I_Ec} as
\begin{eqnarray}\label{intI}
 \mathcal M(s_E)&=&e^2V^{\mu} V^{*\nu}  \times \int \frac{d^4 q_E}{(2\pi)^4}
\left(\sum _{\lambda}\frac{ \left(  \varepsilon^{(\lambda)}_{E \mu }  
   \varepsilon^{*(\lambda)}_{E \nu }   \left(-1+
   i\lambda \beta \sqrt{\gamma  q_E^2}\right) \right)_{q_{4}=s_4} \delta(q_4-s_4 )   }{
 (q_E^2-i\epsilon)  (1+ \beta^2 \gamma q_E^2-i\epsilon)   }  \right)
\nonumber\\ &\times& 
  \delta^{(3)}(\vec q-\vec  p-\vec  p') \,.
\end{eqnarray}
Also, let us write the denominator in \eqref{intI} as
\begin{eqnarray}\label{deno}
 \frac{m^2_{\Lambda}}{(q_E^2-i\epsilon)  (m^2_{\Lambda}+  q_E^2 -i\epsilon )} = \frac{1 }{(q_E^2-i\epsilon)  } -
  \frac{1}{  (  q_E^2+m^2_{\Lambda} -i\epsilon )}    \,,
\end{eqnarray}
where $(\beta^2 \gamma)^{-1}=m^2_{\Lambda}$. We identify two solutions
\begin{eqnarray}
  \omega  =|\vec q|\,, \qquad 
W=\sqrt{ |\vec q|^2+m^2_{\Lambda} }\,,
\end{eqnarray}
where the first one is the standard photon solution and the second one, which arises at
a higher scale $m_\Lambda \sim M$ associated to a massive ghost.

To compute the discontinuities in terms of $s_4$,
 we focus on the element  
\begin{eqnarray}
F(s_4)= \frac{1 }{(q_E^2-i\epsilon)  } -
  \frac{1}{  (  q_E^2+m^2_{\Lambda} -i\epsilon )}    \,,
\end{eqnarray}
and rewrite as 
\begin{eqnarray}
F(s_4)= \frac{1 }{(s_4^2+\omega^2-i\epsilon)  } -
  \frac{1}{  (  s_4^2+W^2 -i\epsilon )}    \,.
\end{eqnarray}
Now we decompose each term as 
\begin{eqnarray}
 \frac{1 }{(s_4^2+x^2-i\epsilon)  } & =& \frac{1}{2ix} \left [  \frac{1 }{(s_4-i x-\epsilon)  } -
  \frac{1}{  (  s_4+ix +\epsilon )} \right]  \nonumber \\  &=& \frac{1}{2x} \left [  \frac{1 }{(i s_4+x-i \epsilon)  } -
  \frac{1}{  (  i s_4-x +i\epsilon )} \right]    \,,
\end{eqnarray}
where some $\epsilon$ terms have been neglected in the numerator. 
Next, we introduce the extended delta to complex variables
\begin{eqnarray}\label{P-V}
\lim_{ \epsilon \to 0 }   \left[  \frac{1}{z -i\epsilon}- \frac{1}{z+ i\epsilon} \right] =2 i\pi \tilde \delta(z)\,,
\end{eqnarray}
meaning that it vanishes everywhere except at some value at the real axis, where it
 reduces to the standard delta function~\cite{Piva-Anselmi}.
Applied to our case we arrive at
\begin{eqnarray}\label{effect}
\text {Disc}\,[F(s_4)]&=& 
\frac{2i\pi \Big(\tilde  \delta( is_4+ \omega )+ \tilde \delta( is_4-\omega ) \Big)   }{  
   2  \omega  }   -
\frac{2i\pi \Big(  \tilde \delta( is_4+W )+ \tilde \delta( is_4-W ) \Big)   }{  
   2W     }  \,.
\end{eqnarray}
Since at effective energies
the external momenta is always much less than
the high energy scale defined by $M$ we set 
 the two last delta functions to zero.  From the expression (\ref{effect}) one has
\begin{eqnarray}
\text {Disc}\,[F(s_4)]= 
\frac{2i\pi \Big( \tilde \delta( is_4+\omega )+ \tilde \delta( is_4-\omega ) \Big)   }{  
   2\omega    } \,.
\end{eqnarray}
Substituting this expression into Eq.~\eqref{intI}, we find 
\begin{eqnarray}
\text {Disc}\, [   \mathcal M(s_E) ]&=&e^2V^{\mu} V^{*\nu}  \times \int \frac{d^4 q_E}{(2\pi)^4}
\left(\sum _{\lambda} \frac{ \left(  \varepsilon^{(\lambda)}_{E \mu }  
   \varepsilon^{*(\lambda)}_{E \nu }   \left(-1+i\lambda \beta \sqrt{\gamma  
   q_E^2} \right) \right)_{q_4=s_4} }  { \beta^2 \gamma}
\nonumber   \right. \\ &\times& \left. 
\frac{(-2i\pi)\Big(  \tilde \delta( i s_4+\omega )+  \tilde \delta(i s_4-\omega ) \Big)   }{  
   2\omega  (\omega -W) (\omega +W)   } \right)
 \delta(q_4-s_4 )  \delta^{(3)}(\vec q-\vec  p-\vec  p') \,.
\end{eqnarray}
Now, evaluating conveniently the delta functions we write
\begin{eqnarray}
\text {Disc}\,[ \mathcal M(s_E)]&=& - e^2V^{\mu} V^{*\nu}  \times\int \frac{d^4 q_E}{(2\pi)^4}
\sum _{\lambda}    \left (  \varepsilon^{(\lambda)}_{E \mu }  
   \varepsilon^{*(\lambda)}_{E \nu } 
  \right)_{q_{4}=s_4}    (2\pi)  \delta(q_4-s_4 ) \delta^{(3)}(\vec q-\vec  p-\vec  p')  
     \\ &\times&    \left[ \frac{ \tilde  \delta( iq_{4}+\omega ) \left(-1+i\lambda \beta \sqrt{\gamma  q_E^2}\right)
    _{q_{4}=i\omega}  }{  \beta^2 \gamma  (q_{4}+i\omega ) 
    (q_{4} -iW)  (q_{4} +iW)   }+ \frac{ \tilde  \delta( i q_{4}-\omega )\left(-1+i\lambda \beta \sqrt{\gamma  
   q_E^2}\right) _{q_{4}=-i\omega} }{\beta^2 \gamma  (-q_{4}+i\omega ) (q_{4} -iW)  (q_{4} +iW) } \right] \nonumber  \,.
\end{eqnarray}
We can obtain this expression in an equivalent way by introducing 
a physical delta function $\bar \delta$
 defined to select only asymptotic degrees of freedom in Hilbert space.
In~\cite{pert_scalar} it has been used to test unitarity in a higher-order Lorentz violating scalar theory. 

The square parenthesis above can be written as
\begin{eqnarray}
\Big[\theta (is_{4})  +\theta (-is_{4})\Big] \bar  \delta(-q_E^2-a_E\sqrt{D_E})=\frac{\theta (is_{4})
  (-1+i\lambda \beta \sqrt{\gamma  q_E^2}) _{{q_{4}}=i\omega} \tilde \delta( iq_{4}+\omega )
 }{\beta^2 \gamma   (q_{4}+i\omega) (q_{4}-iW) (q_{4}+iW) } \nonumber \\ +\frac{\theta (-is_{4})
 (-1+i\lambda \beta \sqrt{\gamma  q_E^2}) _{q_{4}=-i\omega} \tilde \delta( iq_{4}-\omega )
 }{\beta^2 \gamma   (-q_{4}+i\omega) (q_{4}-iW) (q_{4}+iW) }\,,
\end{eqnarray}
where one has to restrict to purely imaginary values of $s_4$, which is precisely the 
case we seek to perform the inverse transformation
of time variable. This allows us to write 
\begin{eqnarray}
\text {Disc}\, [\mathcal M(s_E)]&=&-e^2\int \frac{d^4 q_E}{(2\pi)^4}
  \sum _{\lambda} (     V^{\mu}  \varepsilon^{(\lambda)}_{E \mu } )   
    ( V^{*\nu}\varepsilon^{*(\lambda)}_{ E\nu }    )   \Big[\theta ({is_{4}})  +\theta (-{is_{4}})\Big] (2\pi) \bar
      \delta(-q_E^2-a_E\sqrt{D_E}) \nonumber  \\ &\times&\delta^{(4)}( q_E-  p_E-  p'_E) \,.
\end{eqnarray}
Now, we integrate and consider the inverse transformation of external momenta in terms of $s_0$ 
and use $\text {Disc}\, \mathcal [M]=2i  \text {Im}\, \mathcal M$, to arrive at
\begin{eqnarray}
2  \text {Im}\, \mathcal M(s)= \int \frac{d^4 q}{(2\pi)^4}
  \sum _{\lambda}  |M_{\lambda}|^2 \Big[     \theta (q_{0})  + \theta (-q_{0})\Big]  (2\pi) \bar  \delta(q^2-a\sqrt{D})   \delta^{(4)}( q-  p-  p') \,,
\end{eqnarray}
where 
\begin{eqnarray}
M_{\lambda}=  (-ie)  V^{\mu}  \varepsilon^{(\lambda)}_{ \mu } \,.
\end{eqnarray}
We see that it is equivalent to considering the denominators on-shell in the original expression
or replacing the propagator with the physical delta function.
Therefore the constraint given by unitarity is satisfied.
\subsection{Compton scattering at the one-loop level}
\begin{figure}[h]
\centering
\includegraphics[width=0.5\textwidth]{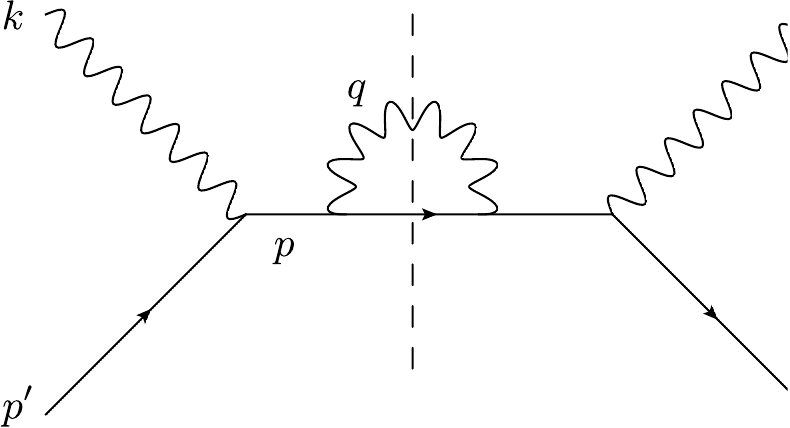}
\caption{\label{Fig2b} The Compton scattering process at one-loop level.}
\end{figure}

Now, we consider the Compton scattering process at the one-loop level. It is presented by Fig.~\ref{Fig2b}.
We set $u^{\mu}=(1,0,0,0)$, so that no ghosts appear. In this case the dispersion relation turn out to be 
\begin{eqnarray}
(q^2)^2 -\frac{16}{M^2}  (q_0^2 (c_1-c_2)+ c_2 | \vec q |^2)  ^2 | \vec q |^2=0 \,,
\end{eqnarray}
Solving, we find the positive solutions 
\begin{eqnarray}\label{solfreq}
\omega_{\lambda'}=\frac{ |\vec q| \sqrt{1-\lambda' g c_2 |\vec q| }    }{\sqrt{1+\lambda' g(c_1-c_2) |\vec q| }} \,,
\end{eqnarray}
where $g=4/M$ and $\lambda'=\pm1$.

The scattering amplitude, with the help of the propagator \eqref{PROPAGATOR},  is found to be
\begin{eqnarray}
i\mathcal M &=& - \sum_{\lambda'} e^4J^{* \mu} (p',k,p)   \int \frac{d^4 q}{(2\pi)^4} 
\frac{(\slashed {p}-\slashed {q} +m)  \varepsilon_{\mu}(q,\lambda')
 \varepsilon^*_{\nu}(q,\lambda')  }{((p-q)^2-m^2+i\epsilon) \left(q^2+\lambda' g \left(q_0^2 (c_1-c_2)  
 +c_2  |\vec q|^2 \right)  |\vec q| +i\epsilon \right) }\nonumber \\&\times&
 J^{\nu}(p',k,p) \,,
\end{eqnarray}
with 
\begin{eqnarray}
J^{\nu}(p',k,p) &=&  
\frac{1}{(p^2-m^2)}  \gamma^{\nu}  (\slashed {p}+m ) \gamma^{\alpha} u^s(p') 
\varepsilon_{\alpha}(k,\lambda)     \;, \nonumber\\
 J^{*\mu}(p',k,p) &=&   
\frac{1}{(p^2-m^2)}\varepsilon^{*}_{\beta}(k,\lambda)  \bar {u}^s(p') 
 \gamma^{\beta}(\slashed {p}+m)
 \gamma^{\mu}  \;.
\end{eqnarray}
We focus on the integral 
\begin{eqnarray}
I_{\mu \nu} (p)&=&   \int \frac{d^4 q}{(2\pi)^4} 
\frac{(\slashed{p}-\slashed{q} +m) \varepsilon_{\mu}(q,\lambda')
 \varepsilon^*_{\nu}(q,\lambda')   }{((p-q)^2-m^2+i\epsilon) \left(q^2+\lambda' g \left(q_0^2 (c_1-c_2)  +c_2  |\vec q|^2 \right)  |\vec q| +i\epsilon \right) }  \,.
\end{eqnarray}
In terms of the poles from Eq. \eqref{solfreq} and the fermion one $E_{q-p}=\sqrt{(\vec q-\vec p)^2-m^2}$, we write
\begin{eqnarray}
I _{\mu \nu}(p)&=&  \int \frac{d^3 qdq_0}{(2\pi)^4} 
\frac{F_{\mu \nu} (p-q,q)}{ (q_0-p_0-E_{q-p}+i\epsilon)  (q_0-p_0+E_{q-p}-i\epsilon)
  }  \nonumber \\ &\times&\frac{1}{(1+\lambda' g (c_1-c_2) |\vec q|     )   (q_0-\omega_{\lambda'}   +i\epsilon) (q_0+\omega_{\lambda'}    -i\epsilon)}\,,
\end{eqnarray}
where 
\begin{eqnarray}
F_{\mu \nu} (p-q,q)&=& (\slashed{p}-\slashed{q} +m) \varepsilon_{\mu}(q,\lambda')
 \varepsilon^*_{\nu}(q,\lambda') \,.
\end{eqnarray}
We perform the $q_0$ integral 
by closing the contour downward and using the residue theorem. Taking into account the relevant poles in
the fourth quadrant, we arrive at
\begin{eqnarray}
 I _{\mu \nu} (p)&=&   \int \frac{d^3 q}{(2\pi)^4}\frac{(-2\pi i)}{(1+\lambda' g (c_1-c_2) |\vec q|     )   } \left[
\frac{[F_{\mu \nu} (p-q,q)]_{q_0=p_0+E_{q-p}-i\epsilon  }}{ 2E_{q-p}  (E_{q-p}+p_0-\omega_{\lambda'}  )
  ( E_{q-p} +p_0 +\omega_{\lambda'} -i\epsilon) }  \right.\nonumber  \\ 
 &-&  \left. \frac{[F_{\mu \nu} (p-q,q)]_{q_0=\omega_{\lambda'}-i\epsilon}}{ 2\omega_{\lambda'}  (E_{q-p}+p_0-\omega_{\lambda'}  )
  ( E_{q-p} -p_0 +\omega_{\lambda'} -i\epsilon) }   \right]\;.
\end{eqnarray}
Using Eq.~(\ref{P-V}), the discontinuity of the integral turns out to be equal to
\begin{eqnarray}
\text{Disc}[\mathcal M(p) ]&=& i \sum_{\lambda'} J^{ \mu} (p',k,p)    Q_{\mu \nu} (p) J^{ \nu} (p',k,p)\,,
\end{eqnarray}
where $Q_{\mu \nu} (p)= \text{Disc} [  I_{\mu \nu}  (p)]$, such that
\begin{eqnarray}
Q_{\mu \nu} (p) &=&  - \int \frac{d^3 q}{(2\pi)^4} \frac{(2\pi)^2}{(1+\lambda' g (c_1-c_2) |\vec q|     )   }\left[
\frac{[F_{\mu \nu} (p-q,q)]_{q_0=p_0+E_{q-p}  }    \delta( E_{q-p} +\omega_{\lambda'}+p_0  )  }
{ 2E_{q-p}  (E_{q-p}+p_0-\omega_{\lambda'}  ) }  \right.\nonumber  \\ &-&
  \left. \frac{[F_{\mu \nu} (p-q,q)]_{q_0=\omega_{\lambda'}}   
  \delta( E_{q-p} +\omega_{\lambda'}-p_0  )   }{2\omega_{\lambda'} (E_{q-p}+p_0-\omega_{\lambda'}  )}   \right]\,.
\end{eqnarray}
We have set $\epsilon=0$ in the numerators 
where the $\epsilon$ factors are not relevant.
Using the delta function, we can simplify the denominators more, i.e.,
\begin{eqnarray}
Q_{\mu \nu} (p) &=&   \int \frac{d^3 q}{(2\pi)^4} \frac{(2\pi)^2}{(1+\lambda' g (c_1-c_2) |\vec q|     )}\left[
\frac{[F_{\mu \nu} (p-q,q)]_{q_0=p_0+E_{q-p}  }    \delta( E_{q-p} +\omega_{\lambda'}+p_0  )  }
{ (2E_{q-p})(  2\omega_{\lambda'} ) }  \right.\nonumber  \\ &+&  \left. \frac{[F_{\mu \nu} (p-q,q)]_{q_0=\omega_{\lambda'}}   
  \delta( E_{q-p} +\omega_{\lambda'}-p_0  )   }{(2E_{q-p})(  2\omega_{\lambda'}  )}   \right]
\end{eqnarray}
With the help of the identity $\int d^3 q=\int d^3 k d^3 k'   \delta^{(3)} (\vec k+\vec k'-\vec p)$,
and introducing two additional integrals in $k_0$ and $k_0'$ 
and with $ k'= p- q$, $ k= q$, we can write
\begin{eqnarray}\label{Qmn}
Q_{\mu \nu} (p)&=&  \int \frac{d^4 k d^4 k'   }{(2\pi)^4}\frac{(2\pi)^2}{(1+\lambda' g (c_1-c_2) |\vec k|     )  }\times\\&\times&
\left[
\frac{[F_{\mu \nu} (p-q,q)]_{q_0=p_0-k'_0 }    \delta(k_0 +k_0'-p_0  )   
 \delta(k_0+\omega_{\lambda'}(k)  )  \delta(k'_0+E_{k'}  )}
{ (2E_{k'})(  2\omega_{\lambda'}(k)  ) }  \right.\nonumber  \\ &+& 
 \left. \frac{[F_{\mu \nu} (p-q,q)]_{q_0=k_0}   
  \delta( k_0 +k_0'-p_0 )  \delta(k_0-\omega_{\lambda'}(k)  )  \delta(k'_0-E_{k'}  )  }
  {(2E_{k'})(  2\omega_{\lambda'}(k)  )}   \right] \delta^{(3)} (\vec k+\vec k'-\vec p)\;.\nonumber
\end{eqnarray}
Now, we use the fact that under the integral with the delta functions, the $F_{\mu\nu}(p-q,q)$ factors behave as
\begin{eqnarray}
[F_{\mu \nu} (p-q,q)]_{q_0=p_0-k'_0 }=[F_{\mu \nu} (p-q,q)]_{q_0=k_0 }   =(\slashed{k}'+m)
 \varepsilon_{\mu}(k,\lambda')
 \varepsilon^*_{\nu}(k,\lambda')   \;,
\end{eqnarray}
and together with the on-shell relation 
\begin{eqnarray}
(\slashed{k}'+m) =\sum_{s'}u^{s'}(k') \bar u^{s'}(k')\;,
\end{eqnarray}
we can rewrite (\ref{Qmn}) as 
\begin{eqnarray}
Q_{\mu \nu} (p)&=&  \sum_{s'} \int \frac{d^4 k   }{(2\pi)^4}  \frac{d^4 k' 
  }{(2\pi)^4}  u^{s'}(k') \bar u^{s'}(k')
\varepsilon_{\mu}  (k,\lambda') \varepsilon^*_{\nu}(k,\lambda')   \times\nonumber\\&\times&
 (2\pi)\delta \left(k^2+\lambda' g \left(k_0^2 (c_1-c_2)  +c_2  |\vec k|^2 \right)  |\vec k|   \right) 
 \nonumber \\ &&(2\pi)\delta(k'^2-m^2 )  \Big[\theta (k_0)\theta (k'_0)
+\theta (-k_0)\theta (-k'_0)  \Big] (2\pi)^4\delta^{(4)} ( k+ k'-p)\,.
\end{eqnarray}
Considering $\text{Disc}[\mathcal M(p) ]=2i\text{Im}[ \mathcal M(p)]$, finally, one has 
\begin{eqnarray}
2\text{Im}[ \mathcal M(p)]&=&\sum_{\lambda',s'} \int  
 \frac{d^4 k   }{(2\pi)^4}  \frac{d^4 k'   }
{(2\pi)^4}  |\tilde M|^2\Big(  \theta (k_0)\theta (k'_0)+    \theta (-k_0)\theta (-k'_0)\Big)\\ &\times& (2\pi)^4\delta^{(4)}
 ( k+ k'-p)    (2\pi)\delta \left(k^2+\lambda' g \left(k_0^2 (c_1-c_2)  +c_2  |\vec k|^2 \right)  |\vec k|   \right) 
 (2\pi)\delta(k'^2-m^2 )\,, \nonumber
\end{eqnarray}
where $\tilde M$ is the diagram obtained by
 replacing the propagators by delta functions after the cutting, i.e.
\begin{eqnarray}
\tilde M = -i e^2 
\frac{1 }{(p^2-m^2)} \varepsilon^*_{\alpha}(k,\lambda') \bar {u}^{s'}(p') 
 \gamma^{\alpha}(\slashed{p}+m) \gamma^{\beta} 
 {u}^{s}(p')  \varepsilon_{\beta}(k,\lambda)   \;.
\end{eqnarray}
Hence we conclude that the optical theorem is satisfied both at the tree level and the one-loop level 
within this scattering process. Since it  is natural to expect that the higher-loop situation does not differ too much,  we conclude that unitarity
is maintained in our theory. 

\section{Summary}\label{V}
 We considered the higher-derivative Lorentz-breaking extension of QED which involves, first, additive terms, that is, Myers-Pospelov and higher-derivative CFJ-like terms, in the purely gauge sector, second, a new, non-renormalizable spinor-vector coupling. For this model, we discussed the dispersion relations and found that, to achieve tree-level unitarity, either only one higher-derivative term, that is, the MP term or the higher-derivative CFJ term can be present in the action, or the Lorentz-breaking vector must be not simply time-like but directed along the time axis.  Apart from this, we carried out study of quantum corrections to two-point functions of gauge and spinor fields and showed that for a consistent subtraction of the divergences, the corresponding higher-derivative terms should be introduced from the very beginning, both in gauge and spinor sectors, with the structure of quantum corrections is simplified for the light-like Lorentz-breaking vectors.  Nevertheless, it is very reasonable to treat this theory as an effective one, aimed for studying of the low-energy domain. Indeed, all higher-order divergent terms will be very small since they are proportional to different degrees of $\frac{1}{M}$, with $M$ is assumed to be of the order of the Planck mass, thus, they are strongly suppressed. One can argue that the similar situation will occur in higher loops where all dangerous divergences will be suppressed by negative degrees of $M$. We carried out a calculation of these corrections in the finite temperature case as well, and we see that our result tends to zero in the high temperature limit.

We verified unitarity in our theory, both at the tree level and at the one-loop level. We checked directly that the optical theorem is satisfied in both cases, therefore, we conclude that, even in the 
presence of higher time derivatives, unitarity in our theory is preserved for the processes we have considered, which rises the hope that other situations, and, in particular, other field theory models, where higher time derivatives do not jeopardize unitarity, are also possible. We conclude that this manner of introducing the higher derivatives is compatible with unitarity as well as the Horava-Lifshitz methodology where only higher spatial derivatives are present. However, the advantage of  our approach is that, unlike the Horava-Lifshitz theories \cite{Horava:2009uw}, in our case the Lorentz symmetry breaking continues to be small which is much more reasonable from the viewpoint of achieving the consistency with experimental measurements, which, as it is well known \cite{datatables}, impose very strong upper boundaries on Lorentz-breaking effects.

{\bf Acknowledgements.} This work was partially supported by Conselho
Nacional de Desenvolvimento Cient\'{\i}fico e Tecnol\'{o}gico (CNPq). The work by A. Yu. P. has been supported by the
CNPq project No. 303783/2015-0.

\end{document}